\documentclass[twocolumn]{autart}    % Enable this line and disable the 
\usepackage{graphicx}          % Include this line if your 
\usepackage{amsmath}
\usepackage{amssymb}
\usepackage{amsfonts}
\usepackage{mathrsfs}

\usepackage{algorithm}
\usepackage{xcolor}
\usepackage{tcolorbox}  
\usepackage{enumitem}
\usepackage{svg}
\usepackage{xpatch}
\usepackage{tikz}
\usepackage{comment}
\usepackage{algorithm}
\usepackage{algorithmic}

\newtheorem{condition}{Condition}

\newtheorem{theorem}{Theorem}

\newtheorem{lemma}{Lemma}

\makeatletter % fix the wrong code in elsart.cls
\xpatchcmd{\runningauthor@fmt}{\global\edef}{\protected@xdef}{}{}
\xpatchcmd{\runningauthor@fmt}{\global\edef}{\protected@xdef}{}{}
\xpatchcmd{\author@fmt}{\edef}{\protected@edef}{}{}
\def\@xnamedef#1{\expandafter\protected@xdef\csname #1\endcsname}
\def\ead@au#1{\protected@edef\@ead@au{#1}}
\def\add@xtok#1#2{\begingroup
  \protected@xdef\@act{\global\noexpand#1{\the#1#2}}\@act
\endgroup}
\def\no@harm{}
\makeatother

\newcommand{\MK}[1]{{\color{black} #1}}

\newcommand{\TR}[1]{{\color{black} #1}}

\newcommand{\xk}{\TR{x_k}}   % change it to x(k) to get back the original notation
\newcommand{\xkp}{\TR{x_{k+1}}}   % change it to x(k+1) to get back the original notation
\newcommand{\uk}{\TR{u_k}}   % change it to u(k) to get back the original notation
\newcommand{\utk}{\TR{u_{k,\theta}}}   % change it to u_\theta(k) to get back the original notation
\newcommand{\vsig}{\TR{z}}
\newcommand{\vsigd}{\TR{\tilde{z}}} %\change this back to \varsigam to get back to the orginal notation

\newcommand{\umstk}{\TR{u_{k,\theta}^\mathrm{ms}}}   % change it to u_\theta^\mathrm{ms}(k) to get back the original notation

\newcommand{\yk}{\TR{y_k}}   % change it to y(k) to get back the original notation

\newcommand{\zk}{\TR{z_k}}   % change it to z(k) to get back the original notation

\begin{document}

\begin{frontmatter}

\title{Gaussian Processes for Optimal Space-Filling Input Design \thanksref{footnoteinfo}} % Title, preferably not more 

\thanks[footnoteinfo]{ Funded by the European Union (ERC, COMPLETE, 101075836). Views and opinions expressed are however those of the author(s) only and do not necessarily reflect those of the European Union or the European Research Council Executive Agency. Neither the European Union nor the granting authority can be held responsible for them.}

\author[Eindhoven]{M{\'a}t{\'e} Kiss}\ead{m.kiss@tue.nl},    % Add the 
\author[Eindhoven]{Yuhan Liu}\ead{y.liu11@tue.nl},               % e-mail address 
\author[Eindhoven]{Maarten Schoukens}\ead{m.schoukens@tue.nl},
\author[Eindhoven,Budapest]{Roland T{\'o}th}\ead{r.toth@tue.nl}% (ead) as shown

\address[Eindhoven]{Control Systems Group, Eindhoven University of Technology, Eindhoven, the Netherlands}
\address[Budapest]{Systems and Control Laboratory, Institute for Computer Science and Control, Budapest Hungary}

\begin{keyword}                           % Five to ten keywords,  
Experiment design; Nonlinear system identification.               % chosen from the IFAC 
\end{keyword}                             % keyword list or with the 
                                          % help of the Automatica 
                                          % keyword wizard

\begin{abstract}                          % Abstract of not more than 200 words.
The quality of a model resulting from (black-box) system identification is highly dependent on the quality of the data that is used during the identification procedure. As nonlinear system identification is sensitive to modeling and extrapolation errors, having experiments that explore the considered operation range of interest are of high importance. The Gaussian-Process based space-filling input design strategy introduces an experiment design technique that aims to achieve a space-filling distribution within the argument space of a (nonlinear) approximator used for modeling the true system. Experiments carried out on simulation examples show that models trained with space-filling signals generalize better over the region of interest compared to models obtained from non-space-filling experiments. Moreover, the method is connected to the optimality theory of experiment design (V- and D-optimal design), allowing the experimenter to tailor the designed space-filling signal to maximize its information content with respect to the experiment objective.
\end{abstract}

\end{frontmatter}

\section{Introduction}
In system identification, we aim to capture a mathematical model of the system behavior from noisy measurement data. The success of such a process hinges on the quality and informativity of the available data. While this is already an important issue for \emph{linear time-invariant} (LTI) system identification, this becomes even more important for nonlinear system identification problems. Indeed, whereas LTI models can be thought of as a hyperplane in a high-dimensional (state) space, nonlinear models are characterized by a manifold \cite{JSchoukens19}, thus being more difficult to extrapolate. In practice, this means that nonlinear models are much more sensitive to modeling errors and assumptions.

Historically, there has been a strong focus on so-called optimal input design, which aims to minimize the variance of the estimated parameters of the identified model, while minimizing the cost of the required experiment, often using a criterion derived from the Fisher information metric. There are state-of-the-art methods for LTI systems based on this concept \cite{annergren2017application,Bombois2021} where the problem effectively translates into a convex optimization. On the other hand, optimal input design for nonlinear system identification is much less understood due to the dependence on time- and frequency-domain characteristics \cite{Cock_D_optimal}. Actual design approaches have been mostly studied in detail for simple nonlinear systems, such as the Hammerstein and Wiener classes \cite{Colin20,colin2020data,Cock_Phd,Forgione2014}.%annergren2017application

However, as argued above, the leading challenge in black-box nonlinear system identification is not to have a small variance on the parameter estimates, but rather to ensure that the model obtains a good data fit over the considered range of operation because the dominant source of errors is that the system is rarely part of the model class. Due to the modeling approximations, there is a discrepancy between the model and the system and we have to shape these modeling errors such that they are small everywhere over the region of operation.
A so called space-filling input is capable to mitigate extrapolation model errors, thus resulting in better generalization over the region of operation \cite{kiss2024space}.

Existing distance-based \cite{herkersdorf2025online,Nelles21,Nelles22} or density-based \cite{herkersdorf2024optimized,vater2024differentiable} methods for such input design objective have shown significant success in increasing accuracy of model estimates. Nevertheless, they are not linked to information theoretic concepts yet and limited to a certain input class, input parametrization or model class and they do not consider input constraints \cite{herkersdorf2024optimized,Nelles21,Nelles22}. In addition, other recent methods \cite{herkersdorf2024optimized,vater2024differentiable} focus on matching the data distribution of a dynamical system with a predefined distribution, which can be restrictive \TR{for real-world} implementation of the resulting signals and increases the complexity of the optimization problem considered.
Active learning methods can also be adopted to explore the argument space through iterative sampling point selection. In these methods, a model is learnt by iteratively optimizing input sequences using variance-based cost \cite{buisson2020actively} or information gain \cite{treven2023optimistic}.

In the preliminary work \cite{liu2025space}, a \emph{Gaussian Process}- (GP) based space-filling input design method for nonlinear system identification has been developed that  allows for a flexible input parameterization, hence offers a control over the amplitude spectrum while respecting constraints on the signals compared to other space-filling input design approaches, as well as applicability to a wider range of nonlinear systems. By placing a \emph{Gaussian Process} (GP) prior on the joint input-state space, the proposed strategy leverages the GP posterior variance to construct a space-filling promoting input design cost function from the Bayesian point of view. This accordingly enables maximization of the coverage in the region of interest, thereby facilitating the generation of informative datasets. Furthermore, \cite{liu2025space} proved that minimization of the mean posterior variance results in a space-filling behaviour.

However, it has not yet been understood how the space-filling input design connects to the optimality theory of experiment design such as A-, D-optimality, etc. While many optimality criteria have a Bayesian equivalent or they can be derived from the appropriate utility function, allowing to incorporate some \emph{a prioiri} information of the system \cite{chaloner1995bayesian,dasgupta1995}, it is largely unexplored how to use such optimality criteria for experiment design for nonlinear system identification. In this work, we establish a GP-based optimal space-filling input design which minimizes the average posterior variance of the function estimate over the design domain, as well as another version that minimizes the Shannon entropy of the distribution of the estimator.
Moreover, we show that the model errors arising from the discrepancies between the system and the assumed model class are reduced with the optimal space-covering input. The space-filling property shapes the model errors such that our estimated model generalizes better over the design domain.

The contributions of this paper are as follows:
%\vspace{-0.8cm}
\begin{enumerate}
    \item Showing the link between GP-based space-filling input design and optimal input design theory, namely %we consider 
    V- and D-optimality.
    \item Demonstrating the effectiveness of the space covering input in achieving accuracy of the model estimates \TR{over a} desired operating range.
\end{enumerate}
The paper is structured as follows. First, Section \ref{sec:ProblemFormulation} describes the problem formulation, including the system and input signal requirements and parametrization. Then, the previously proposed GP-based space-filling input design method is briefly discussed in Section \ref{sec:GPBasedSpaceFillingInputDesign}, followed by proving its link to the optimality theory and introducing a new type of cost function in Section \ref{sec:BayesianOptimalSpaceFillingInputDesign}. To show effectiveness of the experiment design method, simulation results are given in Section \ref{sec:SimulationStudy} and the conclusions on the proposed method are drawn in Section \ref{sec:Conclusion}.

\section{Problem Formulation}\label{sec:ProblemFormulation}
\subsection{Data-Generating Model}
Consider a deterministic system model that, for any specific input $u$,  approximates the \TR{true} process \TR{response with its output $y$}.  %$y$. 
\TR{This model} characterizes our prior knowledge of the system and used to create data distribution inside the argument space of the model. \TR{Accordingly, we} use the following discrete-time state-space representation as a system model, referred to as data-generating model:
\begin{subequations}\label{eq:dataGeneratingSystem}
\begin{align}
    \xkp &= f(\xk, \uk), \label{eq:dataGeneratingSystem_a} \\
    \yk &= g(\xk, \uk), \label{eq:dataGeneratingSystem_b}
\end{align}
\end{subequations}
where $\uk\in\mathcal{U}\subseteq\mathbb{R}^{n_\mathrm{u}}$, $\xk\in\mathcal{X}\subseteq\mathbb{R}^{n_\mathrm{x}}$, and $\yk\in\mathcal{Y}\subseteq\mathbb{R}^{n_\mathrm{y}}$ are the input, state and output signals of the system at time instant $k\in\mathbb{Z}$, respectively, and $\mathcal{Z}\!=\!\mathcal{X}\times\mathcal{U}$ denotes the joint input-state space \TR{with elements $z_k=\mathrm{vec}(x_k,u_k)$}. The smooth nonlinear functions $f(\cdot):\mathcal{Z}\to\mathcal{X}$, $f\in\mathcal{C}_1$ and $g(\cdot):\mathcal{Z}\to\mathcal{Y}$, $g\in\mathcal{C}_1$, \TR{where $\mathcal{C}_1$ denotes continuously differentiable functions,} define the dynamics. \TR{We will require the following condition to be satisfied by \eqref{eq:dataGeneratingSystem}:}
\begin{condition}
    \label{cond:1Controllabilty}
    The state-space representation  \eqref{eq:dataGeneratingSystem} is %assumed to be 
    controllable \TR{in the sense} that there exists a finite time $T_{\mathrm{d}}\in\mathbb{Z}$ that for any two states $x_1,x_2\in\mathcal{X}$ guarantees the existence of an admissible input sequence $u:[0,T_{\mathrm{d}}]\to\mathcal{U}$ that $x_2 = \phi(T_{\mathrm{d}},0,x_1,u)$, where $\phi(t,0,x_0,u)$ is the state transition function corresponding the unique solution of \eqref{eq:dataGeneratingSystem} at time $t\geq0$ for a particular input function $u(\cdot)$ and initial condition $x(0) = x_0$.
\end{condition}
%####################################################################################################################
%####################################################################################################################
%####################################################################################################################
\subsection{Definition of Space-Fillingness}\label{sec:DefinitionOfSpaceFillingness}
\begin{figure}%[htb]
    \centering
    \includegraphics[width=0.9\columnwidth]{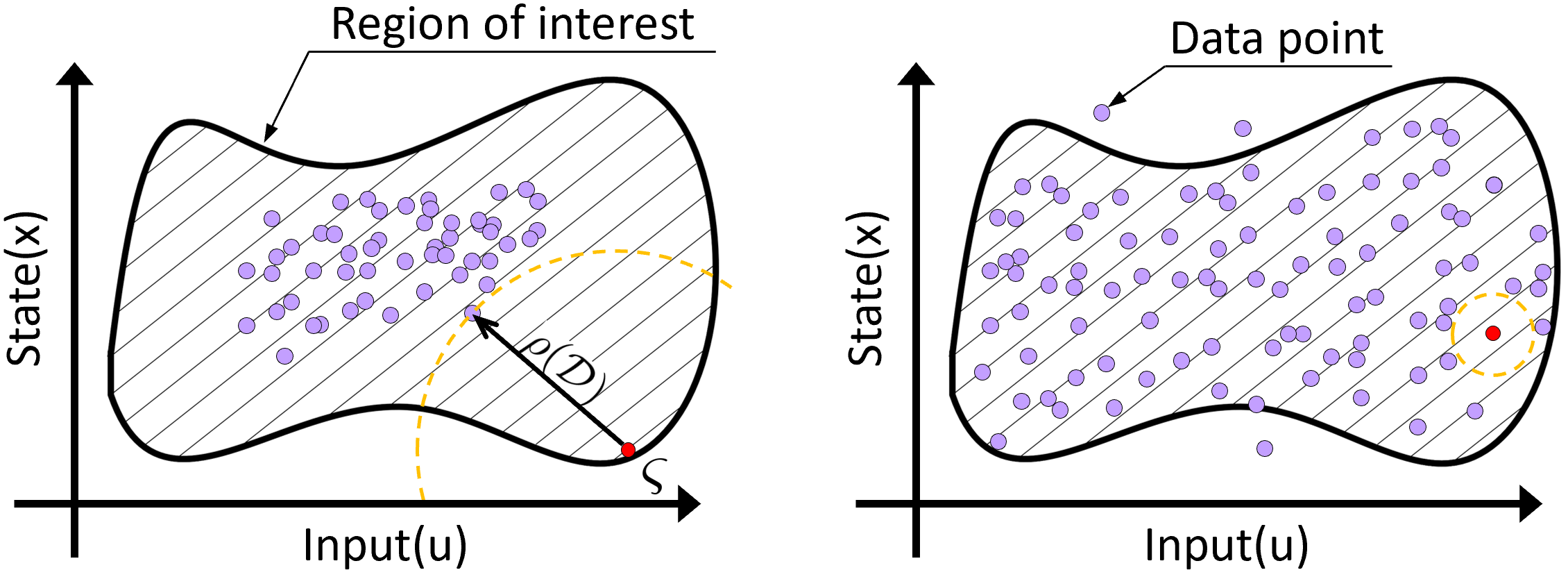}
    \caption{Visualization of the space-filling behavior. Striped domain denotes the region of interest ($\tilde{\mathcal{Z}}$), purple dots denote the dataset ($\mathcal{D}$), yellow circle denotes the largest sphere between the data points with center $\vsigd\!\in\!\tilde{\mathcal{Z}}$.}
    \label{fig:SpaceFillingness}
\end{figure}

In general, we aim in our space-filling design that any point in the space is never too far from a design point $\vsigd$ %($\varsigma$) 
where we intend to evaluate our function \cite{pronzato2012design}. This means, if the model \eqref{eq:dataGeneratingSystem} is excited with an input signal $\{\uk\}_{k=1}^{N}$, the generated data set $\mathcal{D}$ is scattered in the region of interest $\tilde{\mathcal{Z}}$, where the region of interest is a compact subset of the joint input-state space $\tilde{\mathcal{Z}} \subseteq \mathcal{Z}$. Then, we can define the so-called space-filling distance
%-------------------
\begin{equation}
\begin{gathered}
    \rho(\mathcal{D}) = \max_{\vsigd \in \tilde{\mathcal{Z}}}
    \min_{{\vsig} \in \mathcal{D}}
    d(\vsig, \vsigd) ,
\end{gathered}
\end{equation}
where the radius of the largest sphere is $\min_{\vsig \in \mathcal{D}} d(\vsig, \vsigd)$ in terms of a given distance measure $d(\cdot,\cdot):\mathcal{Z}\times\mathcal{Z}\to\mathbb{R}_0^{+}$ (e.g. Euclidean distance) between the data points without containing any other one (see left image in Fig.\ref{fig:SpaceFillingness}).

A data set that minimizes $\rho(\mathcal{D})$ is sought such that the resulting dataset is well scattered inside the region of interest (see right image in Fig.\ref{fig:SpaceFillingness}).  If for a %constant 
\TR{$\epsilon>0$, $\rho({\mathcal{D}})<\epsilon$,} then $\mathcal{D}$ is said to have at least $1/\epsilon$ density w.r.t $\tilde{\mathcal{Z}}$. 

\subsection{Space-Filling Input Design Problem}
In general, %terms, 
space-filling input design aims to find a set of parameters $\theta\!\in\!{\Theta}$ such that it generates a parametrized input sequence $u_{\theta}\!\in\!\mathscr{U}_{\Theta}$. The data set $\mathcal{D}$, which \TR{can be generated by} %results from 
applying this input to the assumed \TR{dynamics in terms of} %known system 
\eqref{eq:dataGeneratingSystem}, should minimize the space-filling metric $\rho(\mathcal{D})$.

Consider the set of parameterized input signals as:
\begin{equation}
   \mathscr{U}_{\Theta} := \{u_\theta : \mathbb{T} \rightarrow \mathbb{R}^{n_{\mathrm{u}}} \mid \theta \in \Theta \}
   \label{eq:inputSignalSet}
\end{equation}
where $\mathbb{T} \subset \mathbb{Z}$ is a finite time \TR{interval. $\mathscr{U}_\Theta$ is assumed to}  %index set and 
satisfy the following conditions:
\begin{condition}
    \label{cond2:Input}
    Each input signal is $u_\theta \in \mathcal{C}^1$ w.r.t. the parameter $\theta\in{\Theta}$ and it is bounded for all $k\in\mathbb{T}$, where ${\Theta}\subseteq\mathbb{R}^{n_\mathrm{\theta}}$ is considered to be a compact set.
\end{condition}
%Furthermore:
\begin{condition}\label{cond:parametrization}
    A sufficiently flexible parametrization exists for $\mathscr{U}_\Theta$ such that for each time step $k$, the input value $\utk$ can be freely specified without constraints by appropriately choosing $\theta$, for example, $\utk\!=\!\theta_k, \forall k\!\in\!\mathbb{T}$ and $\theta\!=\!\mathrm{vec}( \{ \theta_k \}_{k \in \mathbb{T}})$.
\end{condition}

Other than the differentiability and boundedness, there are no more restrictions on the input signals in $\mathscr{U}_{\Theta}$. In other words, (\ref{eq:inputSignalSet}) can represent a wide class of parametric input signals with a broad selection of signal constraints (e.g. spectral constraints). Example of such a parametrization is a multisine signal that excites $L$ number of frequencies with parameters $\theta = \mathrm{vec}(\{A_l,\varphi_l\}_{l=1}^L)$: 
    \begin{equation}
      \umstk \!=\! \sum_{l=1}^{L} A_l \sin\left(2\pi l \frac{\omega_0}{\omega_s}k + \varphi_l\right)
    \label{eq:multisineGeneral}
    \end{equation}
where $u_{\theta}^\mathrm{ms}\!\in\!\mathscr{U}_{\Theta}$ and respectively $\theta\!\in\!\Theta, k\!\in\!\mathbb{T}$. $A_l \in \mathbb{R}^{n_\mathrm{u}}$ denotes the amplitude, $\varphi_l \in [-\pi,\pi]$ denotes the phase, $\omega_0$ corresponds to the frequency resolution and $\omega_s$ denotes the sampling frequency.

To optimally design an input signal under the chosen identification criterion, informativity of the data w.r.t. the chosen model structure is needed \cite{hjalmarsson2021encyclopedia}. However to achieve this, one needs to design an experiment that will generate such a data set, but that can only be done if the underlying system dynamics is perfectly known.
This dependence is a \emph{chicken-egg problem} and leads to a challenge in effectively addressing both the system identification and experiment design task concurrently. To address this difficulty, we formulate the space-filling input design problem based on an approximate system model (\ref{eq:dataGeneratingSystem}), regardless of whether it represents the complete or only the currently available knowledge about the system.
An iterative procedure can be adopted that begins by assuming a region to explore and perform identification based on the dataset collected within this region. Subsequently, using the refined model, the space-filling input design can be repeated, taking into account the previously measured datasets. This alternating procedure continues until it achieves an adequate level of modeling accuracy.

%####################################################################################################################
%####################################################################################################################
%####################################################################################################################
\section{GP-Based Space-filling Input Design}\label{sec:GPBasedSpaceFillingInputDesign}
The mechanism behind our design approach was already detailed in \cite{liu2025space}, where the region of interest was represented with a discrete anchor set. The work proved that if the space-filling cost function $\mathcal{V}(\mathcal{D}_N)$ is zero, it implies that every anchor point has at least one data point on it. Therefore, if the anchor data set $\mathcal{D}_\mathrm{I}$ is designed to be space-filling, the data set $\mathcal{D}_N$ must be also space-filling with at least $1/\epsilon_\mathrm{I}$ density, where $\epsilon_\mathrm{I}$ is the 2-norm distance between the anchor points. This means that the optimal data point distribution will be at least as space-filling as the anchor points (for the full proof see \cite{liu2025space}).

In this section, we deduce how the GP-based space-filling cost function builds up and prove its space-filling behavior, extending the previous work \cite{liu2025space} by incorporating a white noise kernel $\sigma_{\mathrm{e}}^{2} I$. The white noise kernel ensures the full rank property of the resulting posterior covariance matrix and guarantees a lower bound on it which is of importance for the D-optimal analysis later on (see Sec. \ref{sec:DOptSFDesign}). Unlike the earlier approach, this formulation allows for a nonzero cost for the design to be considered space-filling, as long as it is minimal.

\subsection{Preliminaries on Gaussian Process Regression}\label{sec:Preliminaries}
GP regression \cite{rasmussen2003gaussian} is a tool for constructing nonparametric, probabilistic models directly from data and allows for flexible specification of prior assumptions on the system. To construct a \emph{hypothetical} model of
$ f:\mathcal{Z}\to\mathcal{X}$, we consider that candidate estimates $\hat{f}$  of $f$ belong to a Gaussian Process seen as a prior distribution i.e., $\hat{f}\!\sim\!\mathcal{GP} \left( \mu, \kappa \right)$. %with $\hat{f}=\mathrm{vec}(\hat{f}_1,\ldots,\hat{f}_{n_\mathrm{x}})$}. % where $\hat{f}_i\!\in\!\hat{f}\!=\!\mathrm{vec}(\hat{f}_1,\ldots,\hat{f}_{n_x})$.
Consider the dataset $\mathcal{D}_N\!=\!\{Z,F\}$  with the training input matrix $Z =\mathrm{vec}(z_1,...,z_N)\in\mathbb{R}^{\TR{N n_\mathrm{z}}}$ and the training output $F =\mathrm{vec}(f\TR{(z_1)},...,f\TR{(z_N)})\in\mathbb{R}^{ \TR{Nn_\mathrm{x}}}$. %\TR{and $F_{\ell} =\mathrm{vec}(f_{1,\ell},...,f_{N,\ell})\in\mathbb{R}^{N\times 1}$ denoting the outputs for each element $\ell\in\{1,\ldots,n_\mathrm{x}\}$}.
In terms of definition, a \TR{vectorial} \emph{Gaussian Process} (GP) assigns to every point ${z}\!\in\!\mathbb{R}^{n_\mathrm{z}}$ a random variable ${\hat{f}}({z})$ taking values in $\mathbb{R}^{n_\mathrm{x}}$ such that, for any finite set of points $\{{z}_{i}\}_{i=1}^N \subset \mathbb{R}^{n_\mathrm{z}}$, the joint probability distribution of $\hat{f}\left( {z}_1\right), \ldots, \hat{f}\left( {z}_{N}\right)$ is Gaussian. A GP can be fully characterized by its mean function $\mu(z):\mathcal{Z}\to\mathbb{R}^\TR{n_\mathrm{x}}$ and its  \TR{positive-definite} covariance function $\kappa(z,z'):\mathcal{Z}\to\mathbb{R}^\TR{n_\mathrm{x}\times n_\mathrm{x}}$ and it is denoted as: $\hat{f}\sim\mathcal{GP}({\mu}, {\kappa})$.
Then, the joint Gaussian probability of $\hat{f}\left( {z}_1\right), \ldots, \hat{f}\left( {z}_{N}\right)$ is $\mathcal{N}(M_z,K_{Z,Z})$ with $M_z\!=\![\mu(z_1)\ \cdots\ \mu(z_N)]^{\top}$ and $[K_{Z,Z}]_{i,j}\!=\!\kappa(z_i,z_j)$ \TR{block-wise with} $i,j\!\in\!\TR{\mathbb{I}_1^N}=\{1,\ldots,N\}$. In fact, taking $\hat{f} \sim \mathcal{GP} \left( \mu, \kappa \right)$ as the prior distribution in the estimation process defines the prior knowledge about $f$ in terms of the mean function $\mu$, while the choice of $\kappa$ determines the function space in which an estimate of the function is searched for.
Based on the given dataset $\mathcal{D}_N$ and the prior $\hat{f}\!\sim\!\mathcal{GP}(0,\kappa)$ with $\mu\!=\!0$, $\mathbb{P}(F\!\mid\!Z)\!=\!\mathcal{N}(0,K_{Z,Z})$ describes the probability density function of the outputs $F$ seen as random variables conditioned on the observed inputs $Z$.

Given a query point $z_{\ast }\in\mathbb{R}^{n_\mathrm{z}}$, the joint distribution of $F$ and $\hat{f}_{\ast}\!=\!\hat{f}(z_{\ast})$ is:
\begin{equation}
    \left(\begin{array}{c}
    F \\ 
    \hat{f}_{\ast}\end{array} \begin{array}{c}
    \end{array}\right)
    \sim \mathcal{N} \left( 
        0, 
        \begin{bmatrix} 
            K_{Z,Z}+I_{N\TR{n_\mathrm{x}}}\sigma_e^2 & K_Z^{\top}(z_\ast) \\ 
            K_Z(z_\ast) & \kappa(z_{\ast}, z_{\ast}) 
        \end{bmatrix} 
    \right)
\end{equation}
with $[K_Z(z_\ast)]_i\!=\!\kappa(z_i,z_\ast)$ \TR{column blocks}. The Gram matrix $K_{Z,Z}\in\mathbb{R}^{N\TR{n_\mathrm{x}}\times N\TR{n_\mathrm{x}}}$ is  \TR{built-up from $n_\mathrm{x} \times n_\mathrm{x}$ blocks} $[K_{Z,Z}]_{i,j} = \kappa(z_i,z_j)$.

Then, the predictive distribution of $\hat{f}_{\ast}$ given an input $z_{\ast}$ is the posterior
\begin{equation}\label{eq:posteriorDistribution}
    \mathbb{P}(\hat{f}_{\ast} \mid z_{\ast}, \mathcal{D}) = \mathcal{N} \left( \hat{\mu}(\hat{f}_{\ast} \mid z_{\ast}, \mathcal{D}), \hat{c}(\hat{f}_{\ast} \mid z_{\ast}, \mathcal{D}) \right)
\end{equation}
 with the mean expressing the maximum a posteriori estimate (MAP) and variance expressing the uncertainty of the estimate:
\begin{subequations} \label{eq:post}
    \begin{align}
        \label{eq:postMean}\hat{\mu}(\hat{f}_{\ast} \mid z_{\ast}, \mathcal{D})&\!=\!  K_Z(z_\ast)(K_{Z,Z}\!+\!I_{N\TR{n_\mathrm{x}}}\sigma_\mathrm{e}^2)^{-1}F,\\
         \label{eq:postVar}
         \hat{c}(\hat{f}_{\ast} \mid z_{\ast}, \mathcal{D})&\!=\!{\kappa}(z_{\ast},z_{\ast})\! \\
         &-\! K_Z(z_\ast)(K_{Z,Z}\!+\!I_{N\TR{n_\mathrm{x}}}\sigma_\mathrm{e}^2)^{-1} K_Z^{\top}(z_\ast). \notag
     \end{align}
\end{subequations}
The noise variance $\sigma_{\mathrm{e}}^{2}\!\in\!\mathbb{R}^{+}$ is a strictly positive number.
Note that the posterior variance \eqref{eq:postVar} evaluated at $z_{\ast}$ only depends on the GP inputs $Z$, which corresponds to a particular property of the Gaussian process \cite{rasmussen2003gaussian}.

\subsection{Space-filling Promoting Cost Function}\label{sec:Space-filling Promoting Cost Function}
To introduce a space filling cost function for experiment design, we consider a latent \emph{hypothetical} model $\xkp=\hat{f}(\xk,\uk)$ of the data-generating system \eqref{eq:dataGeneratingSystem}, and then we are going to measure the information content of the experiment in terms of the uncertainty of the hypothetical model that would result as the posterior GP estimate based on the given data.
This relies on the data set $\mathcal{D}_N=\{\zk,f(\zk)\}_{k=1}^{N}$, where $\zk = \mathrm{vec}(\xk,\uk)$,   generated by the input sequence $\{ u_{\MK{k},\theta}\}_{k=1}^N$ (we used the shorthand $\uk\!=\!u_{\MK{k},\theta}$) after applying it to the assumed model of the system  \eqref{eq:dataGeneratingSystem}. From this data set $\mathcal{D}_N$, we construct the data \TR{vector} $Z = \mathrm{vec}(\TR{z_1},\ldots,\TR{z_N})\in\mathbb{R}^{\TR{N n_z}}$. Furthermore, we will consider each element $\hat{f}_\ell$ of $\hat{f}=\mathrm{vec}(\hat{f}_1,\ldots,\hat{f}_{n_\mathrm{x}})$ described by an independent Gaussian prior $\hat{f}_{\ell}\!\sim\!\mathcal{GP} \left( 0, \kappa \right)$ sharing the same \TR{scalar} kernel \TR{$\kappa$}. \TR{This corresponds to replacing the matrix kernel in Section \ref{sec:Preliminaries} with $I_{n_\mathrm{x}}\kappa$, where $\kappa(z,z'):\mathcal{Z}\to\mathbb{R}$, essentially  decoupling the posteriori predictive distribution characterized by \eqref{eq:post} to  $\hat{f}_{\ast,\ell} \sim \mathcal{GP}\left({\hat{\mu}_\ell}, \hat{c}\right)$, where
\begin{subequations} \label{eq:post2}
    \begin{align}
        \label{eq:postMean2}\hat{\mu}_{\ell}(\hat{f}_{\ast} \mid z_{\ast}, \mathcal{D})&\!=\!  K_Z(z_\ast)(K_{Z,Z}\!+\!I_N\sigma_\mathrm{e}^2)^{-1}F_\ell,\\
         \label{eq:postVar2}
         \hat{c}(\hat{f}_{\ast} \mid z_{\ast}, \mathcal{D})&\!=\!{\kappa}(z_{\ast},z_{\ast})\! \\
         &-\! K_Z(z_\ast)(K_{Z,Z}\!+\!I_{\MK{N}}\sigma_\mathrm{e}^2)^{-1} K_Z^{\top}(z_\ast). \notag
     \end{align}
\end{subequations}
with $F_\ell$, collecting the observations of $f_\ell$, $K_Z(z_\ast)$ a vector such that $[K_Z(z_\ast)]_i\!=\!\kappa(z_i,z_\ast)$ and $K_{Z,Z}\in\mathbb{R}^{N\times N}$ with elements $[K_{Z,Z}]_{i,j} = \kappa(z_i,z_j)$.}
The \TR{resulting} $\hat{f}_{\ast,\ell}$ belongs to $\hat{f}_\ast\!=\!\mathrm{vec}(\hat{f}_{\ast,\ell},\cdots,\hat{f}_{\ast,n_x})$ where all element \TR{has} identical posterior variance. In the rest of the paper, the dimensional indexing is omitted for the sake of simplicity and we refer to $\hat{f}_\ast$ instead of $\hat{f}_{\ast,\ell}$ unless stated otherwise.

To get a space-filling promoting cost function we need to choose a kernel function that decreases the distance between the data $z$ and a target $z'$ point. Therefore, the following condition has to be satisfied:
\begin{condition}
    \label{cond3:Kernel}
    The kernel function \TR{is composed as} $\kappa(z,z')\!=\!\TR{\bar{\kappa}}(\|z-z'\|)$ \TR{where $\bar{\kappa}$ is a positive function that is monotonically decreasing w.r.t.~the norm $\|z-z'\|$}.
\end{condition}
An appropriate choice is the squared exponential kernel function, i.e.:
\begin{equation}\label{eq:kernel}
    \kappa\left({ {z}},  {{z}}'\right)\!=\!\sigma_{\mathrm{f}}^{2} \exp \left(-\frac{1}{2}( {  {z}}\!-\!  {  {z}}')^{\top} {\Lambda}^{-1}( {  {z}}\!-\! {  {z}}')\right) ,
\end{equation}
with $\sigma_{\mathrm{f}}^{2}\in\mathbb{R}^{+}$, $\Lambda = \mathrm{diag}(\ell_1^2,...,\ell_{n_\mathrm{z}}^2)\in\mathbb{R}^{n_\mathrm{z}\times n_\mathrm{z}}$.

As a next step, we define the cost as the average posterior variance of the model prediction $\hat{f}_{\ast}$ over the design domain $\tilde{\mathcal{Z}}$:
\begin{equation}\label{eq:postVarIntegralForm}
    \mathcal{V}(\mathcal{D}_N) = \int_{\tilde{\mathcal{Z}}} \hat{c}(\hat{f}_{\ast} \mid \tilde{z},\mathcal{D}_N) \hspace{0.3mm} \nu(\tilde{z})\hspace{0.3mm} d\tilde{z},
\end{equation}
where $\nu(\tilde{z})$ is considered as a probability measure over $\tilde{\mathcal{Z}}$ w.r.t.~\TR{a} Lebesgue measure. To avoid the computation of integrals, $\nu(\tilde{z})$ is considered \TR{to correspond to} a uniform distribution and discretized by $M$ number of anchor points. Thus, instead of considering each point of the domain as a query point $z_{\ast}$, the region of interest $\tilde{\mathcal{Z}}$ is described by the anchor dataset $\mathcal{D}_\mathrm{I}:=\{\tilde{z}_i=\mathrm{vec}(\tilde{x}_i,\tilde{u}_i)\}_{i=1}^{M}$ with the anchor \TR{vector} $\tilde{Z}\!=\!\mathrm{vec}(\tilde{z}_1,\cdots,\tilde{z}_\MK{M})\!\in\!\mathbb{R}^{\TR{N n_z}}$ such that they define the $1/\epsilon$ density as well.

\begin{condition}
    \label{cond5:AnchorDataSet}
    The anchor data set is defined as $\mathcal{D}_\mathrm{I}\!:=\!\{\tilde{z}_i=\mathrm{vec}(\tilde{x}_i,\tilde{u}_i)\}_{i=1}^{M}$, such that $\tilde{z}_i\!\neq\!\tilde{z}_j$ for all $i,j\!\in\!\{1,\dots,M\},i\!\neq\!j$, meaning that every anchor point is distinct.
\end{condition}

Since $\mathcal{D}_N$ is generated by a parametrized input $u_{\MK{k},\theta}$, and applied to the system \eqref{eq:dataGeneratingSystem}, resulting in $\mathcal{D}_N(\theta)$, the posterior variance at each anchor point can be expressed as a function of $\theta$:
\begin{align}
   \hat{c}(\hat{f}_{\ast} \mid \tilde{z}_i, \mathcal{D}_N(\theta)) &= \kappa(\tilde{z}_i, \tilde{z}_i) \\
   &-\! \TR{K_{Z_{\theta}}}(\tilde{z}_i) (K_{Z_\TR{\theta},Z_\TR{\theta}}\!+\!I_\TR{N}\sigma_{\mathrm{e}}^{2})^{-1} \TR{K_{Z_{\theta}}^\top}(\tilde{z}_i),\notag
\end{align}
with $\sigma_e^2$ the white noise variance. %\MK{to ensure the positive definiteness of the resulting posterior covariance matrix}
The scalar valued space-filling cost function, namely the average posterior variance over the region of interest evaluated at the anchor points, is
\begin{equation}\label{eq:costFunction}
    \mathcal{V}(\theta):=\frac{1}{M} \sum_{i=1}^{M} \hat{c}(\hat{f}_{\ast} \mid \tilde{z}_i,\mathcal{D}_N(\theta))
\end{equation}
and the space-filling input design problem can be expressed as an optimization problem to find a parameter vector $\theta$ that yields an input sequence $u_{\MK{k},\theta}$ capable of generating a space-filling dataset $\mathcal{D}_N(\theta)$ in case of applying to the system \eqref{eq:dataGeneratingSystem} with \TR{a} given initial condition \TR{$x_0$,} %,u_0$, 
such that the resulting variance $\eqref{eq:costFunction}$ is minimal, ensuring the space-filling behavior within the region of interest $\tilde{\mathcal{Z}}$:
%-------------------------------
\begin{subequations}\label{eq:VOptOptimizationProblem}
\begin{align}
\underset{\theta\in{\Theta}}{\text{min}} 
 & \quad \mathcal{V}({\theta}) \label{f1}\\
 \text{s.t.} & \quad x_\MK{0}=x_\MK{\circ}, %u_\TR{\theta}(0)=u_0},
 \\
 & \quad x_{\MK{j+1}} \!=\! f(x_\MK{j},{u}_{\MK{j},\TR{\theta}}) \hspace{2.7mm} j\!\in\!\TR{\mathbb{I}_0^{N-1}},\\
 &\quad \mathcal{D}_N(\theta)\!=\!\{z_j\!=\!\mathrm{vec}(x_\MK{j},u_{\MK{j},\TR{\theta}}\}_{j=1}^{N}.
\end{align}
\end{subequations}
%------------------------------------

The proposed method is summarized in Algorithm \ref{alg:1}.
\begin{algorithm}%[tb]
   \caption{GP-based Space-Filling Input Design}
   \label{alg:1}
\begin{algorithmic}[1]
   \REQUIRE Select the number of samples $N$, \TR{initial condition $x_0$}, step size $\alpha$, threshold $\gamma$ and the region of interest $\tilde{\mathcal{Z}}$ together with the anchor dataset $\mathcal{D}_\mathrm{I}$.  Set the GP kernel $\kappa(z,z')$ with pre-defined hyperparameters $\Lambda$, $\sigma_{\mathrm{f}}^{2}$, $\sigma_e^2$.
   \STATE  {\bf set} iteration index $\TR{\tau}\!\leftarrow\!0$ 
   and %$u(k,\theta^{(0)})$ with 
   $\theta^{(0)} \leftarrow \theta_0$
  % \STATE{\bf{Step 2:}}
   \REPEAT
    \STATE {\bf simulate}  \eqref{eq:dataGeneratingSystem} with $u_{\theta^{(\TR{\tau})}}$  and obtain $\mathcal{D}_N(\theta^{(\TR{\tau})})$;
    \STATE {\bf compute} $\mathcal{V}(\theta^{(\TR{\tau})})$ according to \eqref{eq:costFunction};
    \STATE {\bf set} $\theta^{(\TR{\tau}+1)}=\theta^{(\TR{\tau})}-\alpha\nabla_{\theta}\mathcal{V}(\theta^{(\TR{\tau})})$;
   % Solve $\theta^{(\TR{\tau}+1)}=\mathrm{argmin}_{ \theta\in\Theta}~ \mathcal{W}(\theta^{(\TR{\tau})};\mathcal{D}_N)$;
   \STATE {\bf set} $\TR{\tau}\gets\TR{\tau}+1$;
    \UNTIL{$\|\theta^{(\TR{\tau})} - \theta^{(\TR{\tau}-1)}\|\leq\gamma$}
   \RETURN $\theta^{(\TR{\tau})}$ and $u_{\theta^{(\TR{\tau})}}(k)$ as the space-filling input.
\end{algorithmic}
\end{algorithm}
\subsection{Theoretical analysis}\label{sec:TheoryVopt}
Based on the results in \cite{liu2025space}, we state the following theorem to show space-fillingness of the designed input signal via the optimization problem \eqref{eq:VOptOptimizationProblem}:
\begin{theorem}\label{theorem1}
     Consider the controllable system \eqref{eq:dataGeneratingSystem}, a parametrized input signal $u_{\MK{k},\theta}$ which satisfies Condition~\ref{cond2:Input} and Condition~\ref{cond:parametrization}. Construct the cost function \eqref{eq:costFunction} with the kernel \eqref{eq:kernel}, satisfying Condition~\ref{cond3:Kernel}, over the discretized form of the region of interest $\tilde{\mathcal{Z}}$ in terms of the $\epsilon$-space filling anchor set $\mathcal{D}_\mathrm{I}$ fulfilling Condition~\ref{cond5:AnchorDataSet}. Design the optimal space-filling input signal based on \eqref{eq:costFunction}. Then, the data set $\mathcal{D}_N(\hat{\theta})$ generated by the optimized input signal $u_{\MK{k},\hat{\theta}}$ applied on \eqref{eq:dataGeneratingSystem}, achieves $\epsilon$-space filling property in the region of interest $\tilde{\mathcal{Z}}$ with a finite data length $N\!\geq\!M\!\times\!Td\!+\!M\!-\!1$, where $\hat{\theta}=\mathrm{arg\,min}_{\theta\in\Theta}~ \mathcal{V}(\theta)$.
\end{theorem}

The first step to prove the theorem is to show the space-filling property of the designed data set $\mathcal{D}_N$ w.r.t. a space-filling anchor data set $\mathcal{D}_\mathrm{I}$. This leads to the following lemma:\\
\begin{lemma}\label{lemma1}
    Given a data set \MK{$\mathcal{D}_N=\{z_k,f(z_k)\}_{k=1}^{N}$} and a region of interest $\tilde{\mathcal{Z}}$ which is characterized by $M$ predefined $\epsilon$-space filling anchor points $\mathcal{D}_\mathrm{I}:=\{\tilde{z}_i\}_{i=1}^{M}$. The GP-based space-filling promoting cost function is defined as \eqref{eq:costFunction}, with the kernel function \eqref{eq:kernel} satisfying Condition~\ref{cond3:Kernel}. If:
\begin{enumerate}[label=\roman*.]
  \item \label{it:1V} the posterior variance $\hat{c}(\tilde{z}_i)$ attains its minimal value if and only if there is a data point on the anchor point;
  \item \label{it:2V}$\mathcal{V}(\theta)$ attains its minimal value if and only if there is at least a data point on every anchor point;
\end{enumerate}
are true (where we used the shorthand notation $\hat{c}(\MK{\hat{f}}_{\ast}\!\mid\!\tilde{z}_i,D_N(\MK{\theta}))\!=\!\hat{c}(\tilde{z}_i)$), then $\mathcal{D}_N$ is guaranteed to have at least $1/\epsilon$ density w.r.t.~$\tilde{\mathcal{Z}}$.
\end{lemma}

\textbf{Proof of~\ref{it:1V}:} We will prove in this point that the posterior variance at $\tilde{z}_i$ is given as \MK{$\hat{c}(\tilde{z}_i)\!=\!\kappa(\tilde{z_i}, \tilde{z_i})\!-\!K_{Z_\theta}(\tilde{z_i}) (K_{Z_\theta,Z_\theta}+I_N\sigma_e^2)^{-1} K_{Z_\theta}^\top(\tilde{z_i})$} attains its minimal value \emph{if and only if} there exists a $j\in\mathbb{I}_1^{N}$ such that  $\tilde{z}_i\!=\!z^{(j)}$.\\
(\emph{Sufficiency})
Assume that $\tilde{z}_i\!\in\!\mathcal{D}_N$, i.e., $\exists j\!\in\!\mathbb{I}_1^{N}$ such that $\tilde{z}_i\!=\!z^{(j)}$. This implies that \MK{$K_{Z_\theta}(\tilde{z_i})=K_{Z_\theta}(z^{(j)})$} is the $j^{\mathrm{th}}$ row of the Gram matrix \MK{$K_{Z_\theta,Z_\theta}$}. If $\mathrm{e}_j\!=\!( \delta_{1,j}, \dots, \delta_{N,j} )$ is a standard basis of $\mathbb{R}^N$ with $\delta_{j,i}\!=\!1$ if  $j\!=\!i$,  and  $\delta_{j,i}\!=\!0$ if  $j\!\neq\!i$, then it is true that $\MK{K_{Z_\theta}(\tilde{z_i})}\!=\!\mathrm{e}_j\MK{K_{Z_\theta,Z_\theta}}$. Substituting this into the posterior covariance formula gives \MK{$\hat{c}(\tilde{z}_i) = \mathrm{e}_j K_{Z_\theta,Z_\theta} \mathrm{e}_j^\top\!-\!\mathrm{e}_j K_{Z_\theta,Z_\theta} (K_{Z_\theta,Z_\theta}\!+\!I\sigma_\mathrm{e}^2)^{-1}K_{Z_\theta,Z_\theta}^{\top}\mathrm{e}_j^{\top}$}.

It is straightforward to see that the quadratic term of the kernel vanishes at \MK{$\kappa(\tilde{z_i},\tilde{z_i})$}, consequently $\mathrm{e}_j \MK{K_{Z_\theta,Z_\theta}} \mathrm{e}_j^\top\!=\!\sigma_\mathrm{f}^2$.
Then, the variance reaches its minimum possible value
\begin{equation}\label{eq:Vproof1}
    \hat{c}(\tilde{z}_i) = \sigma_\mathrm{f}^2 - Q,
\end{equation}
where \MK{$Q\!=\!\mathrm{e}_j K_{Z_\theta,Z_\theta} (K_{Z_\theta,Z_\theta}\!+\!I_N\sigma_\mathrm{e}^2)^{-1}K_{Z_\theta,Z_\theta}^{\top}\mathrm{e}_j^{\top}$} is the posterior reduction term and
\begin{equation}\label{eq:Vproof2}
    \MK{\mathrm{e}_jK_{Z_\theta,Z_\theta}\!=\![\kappa(z^{(j)},z^{(1)}), \ldots, \sigma_{\mathrm{f}}^2, \ldots, \kappa(z^{(j)},z^{(N)})]}
\end{equation}
where the $jj$-th element is $\kappa(z^{(j)},z^{(j)})\!=\!\sigma_{\mathrm{f}}^2$.\\
(\emph{Necessity})
Suppose that $\tilde{z}_i \notin \mathcal{D}_N$ so $\tilde{z}_i\!\neq\!z^{(j)}$. The posterior variance is given by
\begin{equation}\label{eq:Vproof3}
    \hat{c}(\tilde{z}_i) = \sigma_\mathrm{f}^2 - R,
\end{equation}\label{eq:Vproof4}
where \MK{$R\!=\!K_{Z_\theta}(\tilde{z_i}) (K_{Z_\theta,Z_\theta}+I_N\sigma_e^2)^{-1} K_{Z_\theta}^\top(\tilde{z_i})$} and
\begin{equation}
    K_{Z_\theta}(\tilde{z_i})\!=\![\kappa(\tilde{z}_i,z^{(1)}), \ldots, \kappa(\tilde{z}_i,z^{(N)})] .
\end{equation}
We aim to show that $R\!<\!Q$ as a next step, where we define \MK{$K_\mathrm{e}\!=\!K_{Z_\theta,Z_\theta}\!+\!I_N\sigma_e^2$}, $v\!=\!e_j\MK{K_{Z_\theta,Z_\theta}}$, $\tilde{v}\!=\!\MK{K_{Z_\theta}(\tilde{z_i})}$. With this notation, the inequality can be rewritten as $R\!=\!\tilde{v}K_\mathrm{e}^{-1}\tilde{v}^\top\!<\!{v}K_\mathrm{e}^{-1}{v}^\top\!=\!Q$.
The kernel function is monotonically decreasing w.r.t. distance metric $\|z-z'\|$ (see Eq.~\eqref{eq:kernel}), thus it has its maximum value $\sigma_{\mathrm{f}}^2$ when $z\!=\!z'$. In particular, $\MK{\kappa(z^{(j)},z^{(j)})}\!=\!\sigma_{\mathrm{f}}^2$ which occurs only if $\tilde{z}_i\!\in\!\mathcal{D}_N$.

Since all entries of $\tilde{v}$ are guaranteed to be smaller than $\sigma_{\mathrm{f}}^2$, whereas $v$ contains at least one element being equal to $\sigma_{\mathrm{f}}^2$. Consequently $\|\tilde{v}\|\!<\!\|{v}\|$. Now, rewrite vector $v\!=\!\alpha\tilde{v}$, $\alpha\!\in\!\mathbb{R}^+$. Then the statement can be rewritten as $R\!=\!\tilde{v}K_\mathrm{e}^{-1}\tilde{v}^\top\!<\!\alpha\tilde{v}K_\mathrm{e}^{-1}\alpha\tilde{v}^\top\!=\!Q$. Now it is obvious that the right hand side is larger than the left one, which means that $R\!<\!Q$. Hence, the posterior reduction term is maximized when $\tilde{z}_i\!\in\!\mathcal{D}_N$. It follows that the posterior variance $\hat{c}(\tilde{z}_i)$ attains its minimum value \eqref{eq:Vproof1} only if there exists at least one data point on the anchor point $\tilde{z}_i\!\in\!\mathcal{D}_N$. This completes the \emph{necessity} part.

\textbf{Proof of~\ref{it:2V}:} Subsequently, we can prove that that the minimum value of $\mathcal{V}(\MK{\theta})\!=\!\frac{1}{M} \sum_{i=1}^{M}\hat{c}(\tilde{z}_i)$ \emph{if and only if} every anchor points has a data point on it.\\
(\emph{Sufficiency})
Assume that $\forall i\!\in\!\mathbb{I}_{i=1}^M$, the anchor point $\tilde{z}_i$ has a data point on it. Based on point \ref{it:1V}, we know that the minimum value of $\hat{c}(\tilde{z}_i)\!=\!\sigma_\mathrm{f}^2\!-\!Q$. By applying the definition of the cost function \eqref{eq:costFunction}, we obtain the minimum $\mathcal{V}(\theta)\!=\!\frac{1}{M} \sum_{i=1}^{M}\hat{c}(\tilde{z}_i)\!=\!\sigma_\mathrm{f}^2\!-\!Q$ in this particular case.\\
(\emph{Necessity})
Suppose that $\mathcal{V}(\theta)=\sigma_\mathrm{f}^2\!-\!Q$. It can be readily shown that $\frac{1}{M} \sum_{i=1}^{M} \hat{c}(\tilde{z}_i)\!=\!\sigma_\mathrm{f}^2\!-\!Q$ since the proof in the first step says that $\hat{c}(\tilde{z}_i)=\sigma_\mathrm{f}^2\!-\!Q$ holds if and only if $\tilde{z}_i\!\in\!\mathcal{D}_N$. Therefore, every anchor point $\tilde{z}_i$ must be a point in $\mathcal{D}_N$, i.e., $\exists \, z^{(j)}\!\in\! Z$  such that  $z^{(j)}\!=\!\tilde{z}_i$. Since the anchor dataset $\mathcal{D}_\mathrm{I}$ is designed to be space-filling inside the region of interest $\tilde{\mathcal{Z}}$, the dataset $\mathcal{D}_N$ must cover $\tilde{\mathcal{Z}}$ as well with the density $1/\epsilon$ defined by the anchor point distribution. This completes the proof.

Now, we have everything to prove Theorem 1, i.e., to show the $\epsilon$-space-filling property of the dataset $\mathcal{D}_N(\hat{\theta})$.\\
\textbf{Proof of Theorem 1:} 
According to the controllability condition (Cond.~\ref{cond:1Controllabilty}) of the model \eqref{eq:dataGeneratingSystem}, starting from $x_0$ and decide on $u_0$ to reach any anchor point, e.g., $\tilde{x}_1, \tilde{u}_1$, in $T_d$ number of time steps. Hence, all $M$ anchor points can be explored with a finite data length $N\!\geq\!M\!\times\!T_d\!+\!M\!-\!1$. According to Cond.~\ref{cond:parametrization}, the input signal $u_{{\theta}}(k)$ has a sufficient parametrization $\theta\!\in\!\Theta$ that can generate a data set $\mathcal{D}_N(\theta)$, after applying it on model \eqref{eq:dataGeneratingSystem}, that satisfies $\mathcal{V}(\theta) = \frac{1}{M} \sum_{i=1}^{M}\hat{c}(\tilde{z}_i)$ (see Lem.~\ref{lemma1}). Accordingly, the optimized data set $\mathcal{D}_N(\hat{\theta})$ generated by $u_{\MK{k},\hat{\theta}}$ is guaranteed to have the $1/\epsilon$ density w.r.t. $\mathcal{\tilde{Z}}$. This completes the proof.

While classical space-filling designs aim to minimize $\rho(\mathcal{D}_N)$, the GP-based approach makes use of the fact that the covariance function inherently defines a distance metric for the data distribution. In this minimax-type of space-filling design (Sec. \ref{sec:DefinitionOfSpaceFillingness}), the decision variable $\theta$ is optimized using a space-filling cost $\mathcal{V}(\theta)$ , which incorporates the covariance function and, in turn, influences $\mathcal{D}_N$ to optimize the space-filling measure $\rho(\mathcal{D}_N)$.

Next, we will show how the space-filling input design problem can be linked to the classical Bayesian optimal experiment design theory.
%####################################################################################################################
%####################################################################################################################
%####################################################################################################################
\section{Bayesian Optimal Space-Filling Input Design}\label{sec:BayesianOptimalSpaceFillingInputDesign}
%This section provides a theoretical proof on how our input design approach can be associated with an optimality criterion. 
First, we briefly introduce what Bayesian experiment design means and then build up the link between our cost function \eqref{eq:costFunction} and the V-optimality criterion.
\subsection{General concept of Bayesian optimality}
Consider the design task $\xi$ of the experiment as choosing the signal parameters $\theta$ such that the generated input signal $u_\theta$ yields to a space-filling data set inside the region of interest $\tilde{\mathcal{Z}}$ when applied to system \eqref{eq:dataGeneratingSystem}. To quantify how good the experiment was, we can consider $\phi(\xi)$ as an information metric or utility that depends on the design task. In general, we aim to select a design $\xi$ that maximizes the information in our experiment
\begin{equation}
    \xi^{\star} = \arg\max_{\xi} \phi(\xi).
\end{equation}
The framework of Bayesian experiment design focuses on choosing an optimal design ($\xi^*$) by using an information-theoretic approach \cite{atkinson1992,chaloner1995bayesian}. It generally refers to any approach that optimizes an objective function that is functional of the posterior of a learnable variable that we would like to get more information about \cite{chaloner1995bayesian,pronzato2013design,rainforth2024modern}. 

In terms of the input design problem considered in this paper, the objective is to choose the signal parameters $\theta$ that maximize the amount of information about the latent model $\hat{f}_\ast \sim \mathcal{GP}\left(\hat{\mu}_\ast, \hat{c}_\ast\right)$. Hence, the problem can be formalized as optimizing the expected utility $U(\mathcal{D}_N(\theta))$ of the collected data $\mathcal{D}_N=\{Z,F\}$.

%####################################################################################################################
%####################################################################################################################
%####################################################################################################################
\subsection{V-Optimal Space-Filling Input Design}
A utility function is reflecting the purpose of the experiment because the utility of the acquired data is measured through the amount of information it provides for an intended objective \cite{lindley1956measure}. Using the function-space formulation of Section \ref{sec:GPBasedSpaceFillingInputDesign}, consider the region of interest $\tilde{\mathcal{Z}}$ as a continuous domain with probability measure $\nu (\tilde{z})$ on it, the random variable $\hat{f}_\ast\!\in\!\MK{\mathcal{F}}$ with the posterior distribution $\mathbb{P}(\hat{f}_{\ast}\!\mid\!\tilde{z}, \mathcal{D}_N)$ coming from \eqref{eq:posteriorDistribution} at the query point $\tilde{z}\!\in\!\tilde{Z}$. The dataset $\mathcal{D}_N\!=\!\{Z,F\}$, where $Z\MK{\in\mathcal{Z}}$ denotes the inputs and $F\MK{\in \mathcal{X}}$ denotes the outputs. 
We use the shorthand notation $\hat{c}(\hat{f}_\ast\!\mid\!\tilde{z},\mathcal{D}_N)\!=\!\hat{c}_\ast$ and $\hat{\mu}(\hat{f}_\ast\!\mid\!\tilde{z},\mathcal{D}_N)\!=\!\hat{\mu}_\ast$ for the posterior variance \eqref{eq:postVar} and mean \eqref{eq:postMean}.

The posterior variance of a GP at  $\tilde{z}$ can be expressed in general as:
\begin{equation}\label{eq:postVarIntVopt}
    \hat{c}_{\ast} = \int_{\MK{\mathcal{F}}} (\hat{f}_{\ast} - \hat{\mu}_\ast)^2 \ \mathbb{P}(\hat{f}_{\ast} \mid \tilde{z}, \mathcal{D}_N) d\hat{f}_{\ast} 
\end{equation}
with $\mathbb{P}(\hat{f}_{\ast}\!\mid\!\tilde{z}, \mathcal{D}_N)$ the posterior distribution of $\hat{f}_{\ast}$. If the user is interested in the average posterior variance of the function estimate $\hat{f}_\ast$ over the design domain $\tilde{\mathcal{Z}}$, then the utility function is of the following form: 
\begin{equation}
    U \left( \mathcal{D} \right) \!=\! \int_{\tilde{\mathcal{Z}}} \int_{\MK{\mathcal{F}}} (\hat{f}_{\ast}\!-\!\hat{\mu}_\ast)^2 \ \mathbb{P}(\hat{f}_{\ast}\!\mid\!\tilde{z}, \mathcal{D}_N) d\hat{f}_{\ast} \nu(\tilde{z}) d \tilde{z} .
    \label{eq:informationMeasure}
\end{equation}
Taking the expectation over the observations $F$ of function \eqref{eq:informationMeasure} results in the so called Bayesian expected V-optimality criterion \cite{atkinson1992,pronzato2013design}:
\begin{equation}\label{eq:V-optCritGeneral}
    \phi_V(Z) = \mathbb{E} \left[ U(\mathcal{D}_N) \right] .
\end{equation}
To show the link between the GP-based space-filling input design and optimal input design theory, we state the following theorem:\\
\begin{theorem}\label{theorem2}
     Consider the data set $\mathcal{D}_N=\{Z,F\}$ obtained by applying the input signal $u_{\MK{k},\theta}$ on system \eqref{eq:dataGeneratingSystem}. The corresponding Gaussian Process posterior variance provided by Eq.~\eqref{eq:postVar} and the space-filling cost function is defined by Eq.~\eqref{eq:costFunction} over the discretized form of the region of interest $\tilde{Z}$ based on the anchor data set $\mathcal{D}_\mathrm{I}:=\{\tilde{z}_i\}_{i=1}^{M}$. Then, under the assumption that $\nu$ is a uniform measure over the design domain, the considered GP-based space-filling cost function \eqref{eq:costFunction} is equal with the V-optimality criterion defined by \eqref{eq:V-optCritGeneral}.
\end{theorem}

\textbf{Proof:}
Since $\mathcal{D}_N=\{Z,F\}$, the utility function 
\begin{equation}\label{eq:VOptProofUtilityOriginalForm}
    U \left( \mathcal{D}_N \right) \!=\! \int_{\tilde{\mathcal{Z}}} \int_{\MK{\mathcal{F}}} (\hat{f}_{\ast}\!-\!\hat{\mu}_\ast)^2 \ \mathbb{P}(\hat{f}_{\ast}\!\mid\!\tilde{z}, \mathcal{D}_N) d\hat{f}_{\ast} \nu(\tilde{z}) d \tilde{z} .
\end{equation}
depends on the observations $F$. Therefore, its average is taken w.r.t. the marginal distribution of $F$ to formulate a design criterion. Hence, the criterion function can be written as:
\begin{equation}\label{eq:VoptProofCriterionFullForm}
    \phi_V(\mathcal{D}_N) =\mathbb{E} \left[ U(Z,F) \right] =  \int_{\MK{\mathcal{X}}} U(Z,F) \mathbb{P}(F\!\mid\!Z) dF
\end{equation}
where $\mathbb{P}(F\!\mid\!Z)$ denotes the marginal distribution of $F$.
As a property of the Gaussian process, the posterior variance becomes independent of the observations \cite{rasmussen2003gaussian}. After expanding \eqref{eq:postVarIntVopt}, it equals with \eqref{eq:postVar}:
\begin{align*}
    \hat{c}_{\ast}(\hat{f}_\ast\!\mid\!\tilde{z},Z) &= \int_{\MK{\mathcal{F}}} (\hat{f}_{\ast} - \hat{\mu}_\ast)^2 \ \mathbb{P}(\hat{f}_{\ast} \mid \tilde{z}, \mathcal{D}_N) d\hat{f}_{\ast} \\
     &= {\kappa}(z_{\ast},z_{\ast}) \\
     &-\MK{K_Z(z_\ast)}(K_{Z,Z}\!+\!I_N\sigma_{\mathrm{e}}^{2})^{-1}\MK{K_Z^{\top}(z_\ast)} ,
\end{align*}
where the target is $z_\ast\!=\!\tilde{z}$. Upon substituting into \eqref{eq:VOptProofUtilityOriginalForm}, the utility function becomes independent of $F$, $U(Z) = \int_{\tilde{\mathcal{Z}}} \hat{c}_{\ast}(f_\ast\!\mid\!\tilde{z},Z) \nu(\tilde{z}) d\tilde{z}$. This allows to factor \MK{the utility function} out from \eqref{eq:VoptProofCriterionFullForm} and to further simplify the criterion such that
\begin{equation}
    \phi_V(\mathcal{D}_N)\!=\!\int_{\tilde{\mathcal{Z}}} \hat{c}_{\ast}\nu(\tilde{z}) d\tilde{z} \int_{\MK{\mathcal{X}}} \mathbb{P}(F\!\mid\!Z) dF.
\end{equation}
As $\int_{\MK{\mathcal{X}}} \mathbb{P}(F\!\mid\!Z)dF\!=\!1$, the criterion function becomes
\begin{equation}
    \phi_V(Z) =  \int_{\tilde{\mathcal{Z}}} \hat{c}_{\ast}(\hat{f}_\ast\!\mid\!\tilde{z},Z) \nu(\tilde{z}) d\tilde{z},
\end{equation}
where the posterior variance $\hat{c}_{\ast}$ comes from equation \eqref{eq:postVar} and $\nu(\tilde{z})$ has a density w.r.t. the Lebesgue measure. This criterion expresses the average posterior variance of the function estimate over the design domain meaning the Bayesian expected V-optimality criterion \cite{atkinson1992}.
With the use of the shorthand notation for the posterior variance, the criterion function $\phi_V$ is equivalent with the equation used for the space-filling cost function \eqref{eq:postVarIntegralForm}.

Numerical calculations can be facilitated by taking $\nu(\tilde{z})$ as a uniform measure over the space and discretize $\tilde{\mathcal{Z}}$ with the anchor dataset $\mathcal{D}_\mathrm{I}\!=\!\{\tilde{z}_i\}_{i=1}^{M}$ as described in Sec. \ref{sec:Space-filling Promoting Cost Function}. By gridding the space with $M$ number of discrete anchor points, the integral term reduces to a sum and $\phi_V$ is obtained as:
\begin{equation}
    \phi_V(\mathcal{D}_N) = \frac{1}{M} \sum_{i=1}^{M} \hat{c}(\hat{f}_{\ast} \mid \tilde{z}_i,\mathcal{D}_N) .
\end{equation}
We emphasize that for notational convenience, $\mathcal{D}_N$ is used as a dependent variable for $\hat{c}_{\ast}$ and $\phi_V$, even though the dependence on the data set is only partial.
The criterion defined by the above equation can be associated with the GP-based space-filling cost function \eqref{eq:costFunction}. Such that it is shown that our space-filling input design procedure is V-optimal and the proof is completed.

\subsection{D-Optimal Space-Filling Input Design}\label{sec:DOptSFDesign}
Classical D-optimality accounts for minimizing the volume of the ellipsoid of concentration of the model parameters \cite{fedorov1972theory}. Due to the nonparametric feature of the GP, the space-filling input design aims to minimize the uncertainty of the function evaluation everywhere over the region of operation. \MK{This uncertainty can also be measured in terms of the D-optimality criterion, which effectively translates to the maximization of the information matrix \cite{pronzato2013design}. Hence, the Bayesian D-optimality criterion is expressed as}
\begin{equation}\label{eq:DoptCrit}
    \phi_D = \text{det} (\mathcal{I}^{-1}),
\end{equation}
where $\mathcal{I}\!\in\!\mathbb{R}^{N \times N}$ is the posterior information matrix. Interested readers may find more information in \cite{lindley1956measure,pronzato2013design}.

Based on the previous paragraph, the Bayesian D-optimal space-filling cost function is given by
\begin{equation}\label{eq:DoptCost}
    \mathcal{V}_{\mathrm{D}}(\theta) = \text{det} (\hat{C}),
\end{equation}
where $\mathcal{I}^{-1}\!=\!\hat{C}\!\in\!\mathbb{R}^{N\times N}$ is the posterior covariance matrix coming from the GP formulation in Sec. \ref{sec:Preliminaries}. The posterior covariance matrix takes the following form:\MK{
\begin{align}\label{eq:PosteriorCovarianceNoisy}
    \hat{C}(\hat{f}_{\ast} \mid \tilde{Z}, D_N) &= K_{\tilde{Z},\tilde{Z}}\\
    &\quad -  K_{Z_\theta,\tilde{Z}} (K_{Z_\theta,Z_\theta}\!+\!I_N\sigma_{\mathrm{e}}^{2})^{-1} K_{Z_\theta,\tilde{Z}}^\top , \notag
\end{align}
where $K_{\tilde{Z},\tilde{Z}}\!\in\!\mathbb{R}^{N\times N}$ with elements $[K_{\tilde{\mathcal{Z}},\tilde{\mathcal{Z}}}]_{i,j}\!=\!\kappa(\tilde{z}_i,\tilde{z}_j)$ and $K_{Z_\theta,\tilde{Z}}$ is defined similarly.}
Consequently, the optimization problem is defined for any given initial condition $x_0,u_0$ as:
\begin{subequations}\label{eq:DoptProblem}
\begin{align}
\underset{\theta\in{\Theta}}{\text{min}} 
 & \quad \mathcal{V}_\mathrm{D}({\theta}) \label{f2}\\
 \text{s.t.} & \quad x_\MK{0}=x_\MK{\circ},\\
 & \quad x_{\MK{j+1}} \!=\! f(x_\MK{j},u_{\MK{j,\theta}})\hspace{2.7mm} j\!\in\!\MK{\mathbb{I}_0^{N-1}},\\
 &\quad \mathcal{D}_N(\theta)\!=\!\{z_j\!=\!\mathrm{vec}(x_\MK{j},u_{\MK{j,\theta}})\}_{j=1}^{N}.
\end{align}
\end{subequations}
We now present the following theorem for space-fillingness:
\begin{theorem}\label{theorem3} 
    Consider the controllable system \eqref{eq:dataGeneratingSystem}, a parametrized input signal $u_{\theta}(k)$ which satisfies Condition~\ref{cond2:Input} and Condition~\ref{cond:parametrization}. Construct the cost function \eqref{eq:DoptCost} with the kernel \eqref{eq:kernel}, satisfying Condition~\ref{cond3:Kernel}, over the discretized region of interest $\tilde{\mathcal{Z}}$ with the $\epsilon$-space filling anchor set $\mathcal{D}_\mathrm{I}$ fulfilling Condition~\ref{cond5:AnchorDataSet}.  Design the optimal space-filling input signal $u_{\hat{\theta}}(k)$ based on \eqref{eq:DoptProblem}. Then, the data set $\mathcal{D}_N(\hat{\theta})$ generated by the optimized input signal, achieves the $\epsilon$-space filling property in the region of interest $\tilde{\mathcal{Z}}$ with a finite data length $N\!\geq\!M\!\times\!Td\!+\!M\!-\!1$, where $\hat{\theta}=\mathrm{arg\,min}_{\theta\in\Theta}~ \mathcal{V}_D(\theta)$.
\end{theorem}

To facilitate the proof of the theorem, we separately show that $\mathcal{D}_{N}$, generated by the proposed cost function \eqref{eq:DoptCost}, is at least as space-filling as the anchor set $\mathcal{D}_I$, thereby the following lemma is stated:
\begin{lemma}
    Given a data set $\mathcal{D}_N\!=\!\{z_j\}_{j=1}^{N}$ and a region of interest $\tilde{\mathcal{Z}}$ which is characterized by $M$ predefined $\epsilon$-space filling anchor points $\mathcal{D}_\mathrm{I}\!=\!\{\tilde{z}_i\}_{i=1}^{M}$ \MK{and the anchor vector $\tilde{Z}\!=\!\{\tilde{z}_i,\cdots,\tilde{z}_M\}$} such that $M\leq N$. Define the GP-based space-filling cost function as \eqref{eq:DoptCost}, with the kernel function \eqref{eq:kernel}.
    The data set $\mathcal{D}_N$ is guaranteed to have at least 1/$\epsilon$ density w.r.t. $\tilde{\mathcal{Z}}$ if and only if there is at least a data point on each anchor point, and consequently $\mathcal{V}_{\mathrm{D}}(\theta)$ becomes minimal.
\end{lemma}

\textbf{Proof:} First, we are going to show that the posterior covariance matrix \MK{$\hat{C}_{\ast}$} is always full rank thus the cost function $\mathcal{V}_D(\mathcal{D}_N)$ is not singular (where we used the shorthand notation for the covariance matrix \MK{$\hat{C}(f_{\ast}\!\mid\!\tilde{Z}, D_N)\!=\!\hat{C}_{\ast}$}).\\
Adopt the notations \MK{$K_{Z,\tilde{Z}}\!=\!K_{Z_\theta,\tilde{Z}}$}, $K_\mathrm{e}\!=\!K_{Z,Z}\!+\!I_N\sigma_\mathrm{e}^2$ and let the posterior covariance matrix be defined as \MK{$\hat{C}_\ast\!=\!K_{\tilde{Z}, \tilde{Z}}\!-\!K_{Z,\tilde{Z}}K_\mathrm{e}^{-1}K_{Z,\tilde{Z}}^{\top}$}.
Consider the augmented kernel matrix $M\!=\!\left[\begin{smallmatrix} K_\mathrm{e} & \MK{K_{Z,\tilde{Z}}^\top} \\ \MK{K_{Z,\tilde{Z}}} & K_{\tilde{Z},\tilde{Z}} \end{smallmatrix}\right]$. The Schur complement of $K_\mathrm{e}$ in $M$ is \MK{$K_{\tilde{Z}, \tilde{Z}}\!-\!K_{Z,\tilde{Z}}K_\mathrm{e}^{-1}K_{Z,\tilde{Z}}^{\top}$} which coincides with the posterior covariance matrix. Given that $\sigma_\mathrm{e}^2\!>\!0$, the matrix $K_\mathrm{e}\!\succ\!0$ and therefore it is invertible. It follows that $M\!\succ\!0$ and \MK{$K_{\tilde{Z}, \tilde{Z}}\!-\!K_{Z,\tilde{Z}}K_\mathrm{e}^{-1}K_{Z,\tilde{Z}}^{\top}\!\succ\!0$} hold, see \cite{boyd2004convex}. This constitutes the positive definiteness of \MK{$\hat{C}_\ast$}, hence it is full rank.

The upcoming point of the proof defines the minimum possible value of the cost and presents that it can be achieved \emph{if and only if} there is at least a data point on each anchor point.\\
(\emph{Sufficiency})
Assume that every anchor point has a data point on it, i.e. $\mathcal{D}_\mathrm{I}\!\subseteq\!\mathcal{D}_N$. This implies that \MK{$K_Z(\tilde{z}_i)\!=\!K_Z(z^{(j)})$} is the $j^{\mathrm{th}}$ row of the Gram matrix $K_{Z,Z}$. The selection matrix $E\!\in\!\mathbb{R}^{M\times N}$, whose $j^{\mathrm{th}}$ row is the standard basis vector $e_{j}$ (see Sec.~\ref{sec:TheoryVopt}), selects the rows of $K_{Z,Z}$ of the coinciding anchor and data points. We may consider the relations $\MK{K_{Z,\tilde{Z}}}\!=\!EK_{Z,Z}$ and $K_{\tilde{Z},\tilde{Z}}\!=\!EK_{Z,Z}E^{\top}$. Substituting these into the formula of the covariance matrix gives $\hat{C}_\ast\!=\!EK_{Z,Z}E^\top\!-\!EK_{Z,Z}(K_{Z,Z}+I_N\sigma_\mathrm{e}^2)^{-1}K_{Z,Z}^\top E^\top$.\\
Consider the Gram matrix as $K_{Z,Z}\!=\!K_{Z,Z}\!+\!I_N\sigma_\mathrm{e}^2\!-\!I_N\sigma_\mathrm{e}^2$. Multiplying both sides with $(K_{Z,Z}\!+\!I\sigma_\mathrm{e}^2)^{-1}$ and simplifying yields:
\begin{equation}\label{eq:DoptProof_relation}
    (K_{Z,Z}\!+\!I\sigma_\mathrm{e}^2)^{-1}K_{Z,Z} = I-\sigma_\mathrm{e}^2(K_{Z,Z}\!+\!I\sigma_\mathrm{e}^2)^{-1}.
\end{equation}
Substituting \eqref{eq:DoptProof_relation} into the posterior covariance expression gives $\MK{\hat{C}_\ast}\!=\!EK_{Z,Z}E^\top\!-\!EK_{Z,Z}(I_N-\sigma_\mathrm{e}^2(K_{Z,Z}\!+\!I_N\sigma_\mathrm{e}^2)^{-1})E^\top$ .
After doing some simplification, the reduced form is
\begin{equation}
    \MK{\hat{C}^{\mathrm{suf}}_\ast}\!=\!\sigma_\mathrm{e}^2EK_{Z,Z}(K_{Z,Z}+I_N\sigma_\mathrm{e}^2)^{-1}E^{\top} .
\end{equation}
Next, the cost function will have the following form
\begin{align}
    \mathcal{V}_\mathrm{D}^{\mathrm{suf}}\!&=\!\text{det}(\hat{C}^{\mathrm{suf}}_\ast)\! \\
    &=\!\mathrm{det}(\sigma_\mathrm{e}^2EK_{Z,Z}(K_{Z,Z}+I_N\sigma_\mathrm{e}^2)^{-1}E^{\top}) \notag
\end{align}
This completes the proof of sufficiency.

(\emph{Necessity}) Now, consider the case that there is at least one anchor point that has no data point on it, i.e. $\mathcal{D}_\mathrm{I}\!\subset\!\mathcal{D}_N$.
The vector of anchor points is partitioned into two parts: coinciding block $Z_c\!\in\!\mathbb{R}^m$ with $\tilde{z}_i\!\in\!Z_c$ if there exist a $j$ when $\tilde{z}_i\!=\!z^{(j)}$; and distinct block $Z_d\!\in\!\mathbb{R}^r$ with $\tilde{z}_i\!\in\!Z_d$ if $\tilde{z}_i\!\neq\!z^{(j)}$. Thus $\MK{\tilde{Z}}\!=\![Z_c,Z_d]^\top\!\in\!\mathbb{R}^M$ where $M\!=\!m\!+\!r$. Consequently, the mixed covariance matrix will have the following structure: $\MK{K_{Z,\tilde{Z}}}\!=\![Z_{c,Z},Z_{d,Z}]^\top\!\in\!\mathbb{R}^{M\times N}$.
The block matrices obtained are: $K_{c,c}\!=\!\kappa(Z_c,Z_c)\!\in\!\mathbb{R}^{m\times m}$, $K_{c,d}\!=\!\kappa(Z_c,Z_d)\!\in\!\mathbb{R}^{m\times r}$, $K_{d,d}\!=\!\kappa(Z_d,Z_d)\!\in\!\mathbb{R}^{r\times r}$, $K_{c,Z}\!=\!\kappa(Z_c,Z)\!\in\!\mathbb{R}^{m\times N}$, $K_{d,Z}\!=\!\kappa(Z_d,Z_Z)\!\in\!\mathbb{R}^{r\times N}$.
That blocks define the following block-structured posterior covariance matrix:
\begin{equation}\label{eq:PostCovMtrxBlockStr}
\begin{split}
    \begin{bmatrix}
    \hat{C}_{cc} & \hat{C}_{cd} \\
    \hat{C}_{cd}^\top & \hat{C}_{dd}
    \end{bmatrix}
    &=
    \begin{bmatrix}
    K_{cc} & K_{cd} \\
    K_{cd}^\top & K_{dd}
    \end{bmatrix} \\
    &-
    \begin{bmatrix}
    K_{c,Z} \\
    K_{d,Z}
    \end{bmatrix}
    (K_{Z,Z}\!+\!I\sigma_\mathrm{e}^2)^{-1}
    \begin{bmatrix}
    K_{c,Z} \\
    K_{d,Z}
    \end{bmatrix}^\top .
\end{split}
\end{equation}
We now proceed by examining each block of \MK{$\hat{C}_\ast$}.
First, we analyze $\hat{C}_{c,c}\!=\!K_{CC}\!\!-\!K_{c,Z}(K_{Z,Z}\!+\!I_N\sigma_\mathrm{e}^2)^{-1}K_{c,Z}^\top$. Let $E_m\!\in\!\mathbb{R}^{m\times N}$ be the selection matrix whose $j^{\mathrm{th}}$ row is the standard basis vector $e_j$ (see Sec.~\ref{sec:TheoryVopt}) selecting the data indices that coincide with the $j^{\mathrm{th}}$ element of $Z_c$. We can write $K_{cZ}\!=\!E_mK_{Z,Z}$ and $K_{cc}\!=\!E_mK_{Z,Z}E_m^\top$. After the substitution into $\hat{C}_{c,c}$ and subsequent simplification, the purely coinciding block of the posterior covariance matrix results in $\hat{C}_{c,c}\!=\!\sigma_\mathrm{e}^2E_mK_{Z,Z}(K_{Z,Z}\!+\!I_N\sigma_\mathrm{e}^2)^{-1}E_m^{\top}$.\\
Moving onto the diagonal element $\hat{C}_{c,d}\!=\!K_{c,d}\!-\!K_{c,Z}(K_{Z,Z}\!+\!I_N\sigma_\mathrm{e}^2)^{-1}K_{d,Z}^\top$, it can be realized that $K_{c,Z}\!=\!E_mK_{Z,Z}$ thus the second term of the expression becomes $E_mK_{Z,Z}(K_{Z,Z}\!+\!I_N\sigma_\mathrm{e}^2)^{-1}K_{d,Z}^\top\!$. Using the relation defined in \eqref{eq:DoptProof_relation}, we observe that the second term becomes $E_mK_{d,Z}^\top\!+\!\sigma_\mathrm{e}^2E_m(K_{Z,Z}\!+\!I_N\sigma_\mathrm{e}^2)^{-1}K_{d,Z}^\top$.
Observe that $E_mK_{d,Z}^\top\!=\!E_mK_{Z,d}\!=\!K_{c,d}$, since $E_m$ selects the coinciding terms of the data $Z$. Hence, the diagonal element $\hat{C}_{c,d}\!=\!\sigma_\mathrm{e}^2E_m(K_{Z,Z}\!+\!I_N\sigma_\mathrm{e}^2)^{-1}K_{d,Z}^\top$.\\
The last block is $\hat{C}_{d,d}\!=\!K_{d,d}\!-\!K_{d,Z}(K_{Z,Z}\!+\!I_N\sigma_\mathrm{e}^2)^{-1}K_{d,Z}^\top$ cannot be further simplified, therefore the resulting posterior covariance matrix will have the following structure
\begin{equation}
\hat{C}^{\mathrm{nec}}_\ast \!=\!
\begin{bmatrix}
    \sigma_\mathrm{e}^2E_mK_{Z,Z}K_{\mathrm{e}}^{-1}E_m^{\top} & [\sigma_\mathrm{e}^2E_mK_{\mathrm{e}}^{-1}K_{d,Z}^\top]^\top \\
    \sigma_\mathrm{e}^2E_mK_{\mathrm{e}}^{-1}K_{d,Z}^\top &
    K_{d,d}\!-\!K_{d,Z} K_{\mathrm{e}}^{-1} K_{d,Z}^\top
\end{bmatrix} ,
\end{equation}
where we used the shorthand notation $K_\mathrm{e}\!=\!K_{Z,Z}\!+\!I_N\sigma_\mathrm{e}^2$.
By computing the determinant of the above expression using the block notations, we obtain that the cost becomes equal to 
\begin{equation}
    \mathcal{V}_\mathrm{D}^{\mathrm{nec}}\!=\!\mathrm{det}(\hat{C}_{cc})\mathrm{det}(\hat{C}_{dd}-\hat{C}_{cd}\hat{C}_{cc}^{-1}\hat{C}_{cd}^\top) .    
\end{equation}

Let the selection matrices be $E_m\!\in\!\mathbb{R}^{m\times N}$ and $E_r\!\in\!\mathbb{R}^{r\times N}$ and define the posterior covariance of the sufficient part as $\MK{\hat{C}^{\mathrm{suf}}_\ast}\!=\!\left[\begin{smallmatrix} \hat{C}^{\mathrm{suf}}_{A} & 0 \\ 0 & \hat{C}^{\mathrm{suf}}_{D} \end{smallmatrix}\right]$, where $\hat{C}^{\mathrm{suf}}_{A}\!=\!\sigma_\mathrm{e}^2E_mK_{Z,Z}(K_{Z,Z}+I_N\sigma_\mathrm{e}^2)^{-1}E_m^{\top}$ and $\hat{C}^{\mathrm{suf}}_{D}\!=\!\sigma_\mathrm{e}^2E_rK_{Z,Z}(K_{Z,Z}+I_N\sigma_\mathrm{e}^2)^{-1}E_r^{\top}$. Now, rewrite the cost function as $\mathcal{V}_\mathrm{D}^{\mathrm{suf}}\!=\!\mathrm{det}(\hat{C}^{\mathrm{suf}}_{A})\mathrm{det}(\hat{C}^{\mathrm{suf}}_{D})$. Since $\mathrm{det}(\hat{C}^{\mathrm{suf}}_{A})\!=\!\mathrm{det}(\hat{C}_{cc})$ and by Theorem~\ref{theorem1}, it follows that $\mathrm{det}(\hat{C}^{\mathrm{suf}}_{D})\!<\!\mathrm{det}(\hat{C}_{dd}-\hat{C}_{cd}\hat{C}_{cc}^{-1}\hat{C}_{cd}^\top)$. Therefore, $\mathcal{V}_\mathrm{D}^{\mathrm{suf}}\!<\!\mathcal{V}_\mathrm{D}^{\mathrm{nec}}$. The D-optimal cost attains its minimum $\mathcal{V}_\mathrm{D}\!=\!\mathrm{det}(\sigma_\mathrm{e}^2EK_{Z,Z}(K_{Z,Z}+I\sigma_\mathrm{e}^2)^{-1}E^{\top})$, \emph{if and only if} there is at least a data point on each anchor point. Therefore, every anchor point $\tilde{z}_i$ must be a point in $\mathcal{D}_N$, i.e., $\exists \, z^{(j)}\!\in\!Z$  such that  $z^{(j)}\!=\!\tilde{z}_i$. As the anchor dataset $\mathcal{D}_\mathrm{I}$ is designed to be space-filling inside the region of interest $\tilde{\mathcal{Z}}$, the dataset $\mathcal{D}_N$ also must cover $\tilde{\mathcal{Z}}$, by that completing the proof.

Theorem 1 has already shown that the optimized data set $\mathcal{D}_N(\hat{\theta})$, generated by the input signal \MK{$u_{k,\hat{\theta}}$} under the space-filling cost function, achieves a $1/\epsilon$ density with respect to $\mathcal{\tilde{Z}}$. Since the only change relative to Theorem 1 is the cost function \eqref{eq:DoptCost}; and from Lemma 2, we establish the $\epsilon$-space-filling property of the optimized data set $\mathcal{D}_N(\hat{\theta})$. It follows directly—without repeating the previous arguments—that the optimized data set $\mathcal{D}_N(\hat{\theta})$ obtained using \eqref{eq:DoptCost}, likewise achieves a $1/\epsilon$ density with respect to $\mathcal{\tilde{Z}}$, thus completing the proof of Theorem 3.

%####################################################################################################################
%####################################################################################################################
%####################################################################################################################
\section{Simulation Study}\label{sec:SimulationStudy}
This section presents a Monte-Carlo study where we first create space-filling signals for 50 different realizations for a mass-spring-damper system and then compare them to unoptimized signals. To quantify the effectiveness of the space-filling design we identify a state-space neural network based on the resulting data set for each realization of the system. We follow the identification method described in \cite{schoukens2021improved} and compare the models based on the model error resulting on a test set coming from the same input class with the same statistics as the initial multisine. The system parameters vary randomly across the realizations and the designs are carried out under the constraint that the energy is limited in the input signal.

\subsection{The Considered System Class}
Assume the data generating model is a second order mass-spring-damper system in the form where the state and output equations can be decomposed into a linear and a nonlinear term:
\begin{equation}
\begin{aligned}
    \xkp &= A\xk+B\uk+\tilde{f}(\xk), \\
    \yk &= C\xk+D\uk,
\end{aligned}
\label{eq:polyNonlinSS}
\end{equation}
where variable $\tilde{f}$ is a static nonlinear function representing a fourth order polynomial nonlinearity $\tilde{f} = \sum_{p=0}^{p=4} a_p x^p$. The coefficients $a_p$ are drawn from the same uniformly distributed random initializations $[-1,1]$. The elements of the state matrices ($A,B,C,D$) are also sampled from different uniform distributions.

\subsection{Input Parametrization}
Each realization of the system is excited by a periodic multisine with $l = 471$ excited frequencies between the range $f_{min}$=1~Hz and $f_{max}$=24~Hz, a sampling frequency $f_s$=50~Hz, and $N$=1024 data samples per period. The signal is parametrized by its phases and amplitude such that
\begin{equation}\label{eq:multisineInit}
    \MK{u_{k,\theta}} = \sum_{l=l_{min}}^{l_{max}} \alpha \sin \left( 2\pi l \frac{f_0}{f_s} k + \varphi_l \right),
\end{equation}
where $f_0 = f_s/N$ and $l_{min} = 22$, $l_{max} = 492$. The multisine is chosen to be zero-mean, the phases are coming from a uniform random initialization $[0,2 \pi)$ and the amplitude has an initial value of 0.2 Newton. The signal is optimized by changing the phases of each frequency component and the amplitude: $\theta=\mathrm{vec}(\alpha, \ \varphi_1 \cdots \varphi_F)$. Note, that in this parametrization every frequency component has the same level of amplitude. The signal constraint bounds $\alpha$ such that the standard deviation of the input is between 0 and 0.4.

\subsection{Region of Interest and Hyper Parameters}
The region of interest is a rectangular box in the 3-dimensional input-state space $\mathcal{\tilde{Z}} \!=\! \{ (u,x_1,x_2) \!\in\! \mathbb{R}^3 \mid u \!\in\! [u_{\min}, u_{\max}], \; x_1 \!\in\! [x_{1,\min}, x_{1,\max}], \; x_2 \!\in\! [x_{2,\min}, x_{2,\max}] \}$; with $M \!=\! 125$, 5 equally distanced anchor points $\tilde{z}_i$ in each dimension with $i\in\mathbb{I}_{1}^{M}$.
The borders of the region of interest are arbitrarily chosen as the range, see Figure \ref{fig:3D} for an example.
In order to have a sufficient overlap between the region of influence of neighboring points, the kernel widths are chosen to be equal with the adjacent anchor point distance in the corresponding dimension. In our 3 dimensional case they are $\Lambda = \mathrm{diag}((\tilde{z}_i^{[1]}-\tilde{z}_{i-1}^{[1]})^2,(\tilde{z}_i^{[2]}-\tilde{z}_{i-1}^{[2]})^2,(\tilde{z}_i^{[3]}-\tilde{z}_{i-1}^{[3]})^2)$ with noise variance $\sigma_e^2\!=\!0.01$ and the scaling factor is $\sigma_{\mathrm{f}}^{2} = \sqrt{10}$.

\subsection{Monte-Carlo Simulation}
We create space-filling signals for 50 different realizations where each realization is starting from a new random phase multisine. One multisine realization is optimized for V-optimality while the other is optimized for D-optimality.  In each Monte Carlo realization, the phases are coming from different uniform random distributions just as the system matrices and polynomial coefficients of \eqref{eq:polyNonlinSS} are assigned to new random parameters while the amplitude and optimization method remains unchanged.

The V-optimal \eqref{eq:costFunction} and D-optimal \eqref{eq:DoptCost} cost functions are nonlinear in the parameters due to (\romannumeral 1) the nonlinear dependence of the input signal on the parameters, (\romannumeral 2) the nonlinear dependence of the state on the (past) inputs and (\romannumeral 3) due to the nonlinear dependence of the cost on the joint input-state-values. A gradient-based solver, Matlab function \verb+fmincon+, is therefore used for the optimization which also allows additional (parameter) constraints to be added during the optimization, e.g. limiting the power at each frequency (in case of a multisine parametrization).

\begin{figure}%[htb]
    \centering
    \includegraphics[width=\columnwidth]{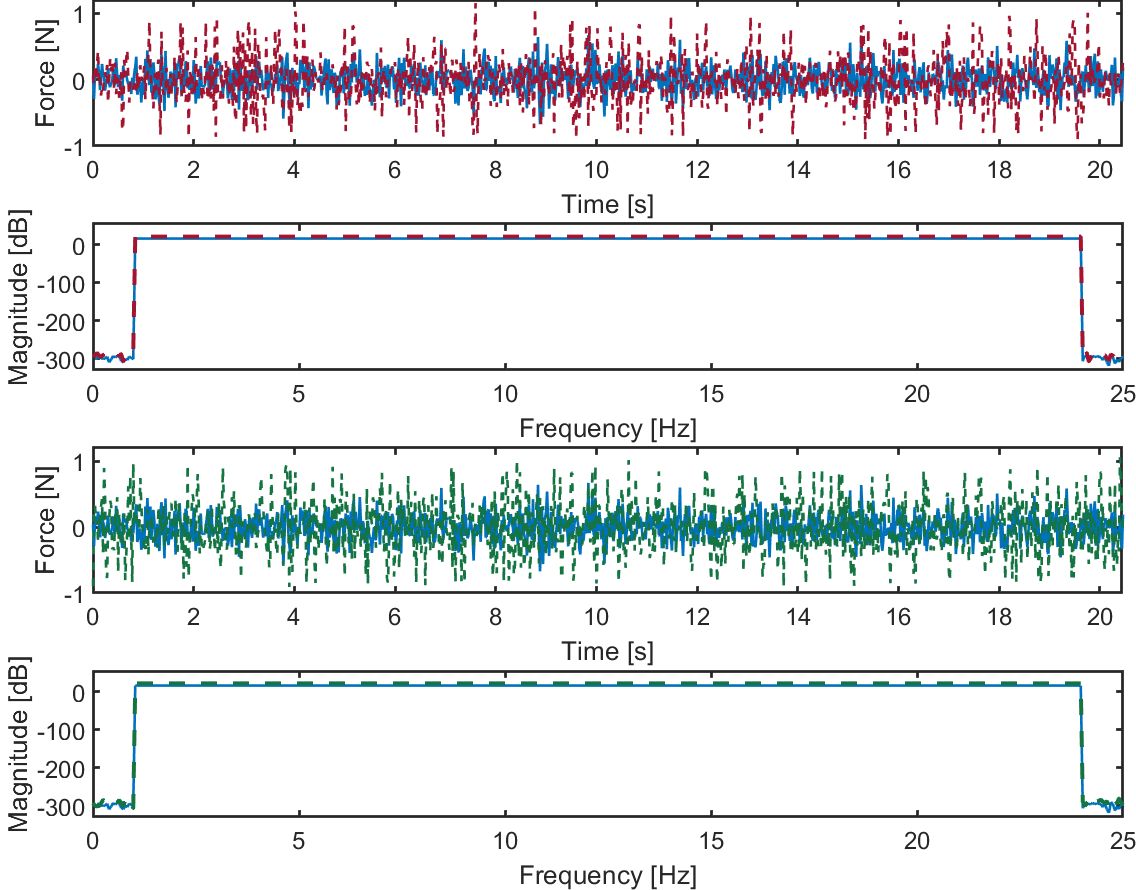}
    \caption{Optimized space-filling signals. Random phase $u_{k,\theta}$ (blue), V-optimal space-filling $u_{k,\hat{\theta}}^V$ (red) and D-optimal space-filling $u_{k,\hat{\theta}}^D$ (green) input in time and frequency domain up to the Nyquist frequency.
    }
    \label{fig:signalComparison}
\end{figure}

\begin{figure}%[htb]
    \centering
    \includegraphics[width=1.0\columnwidth]{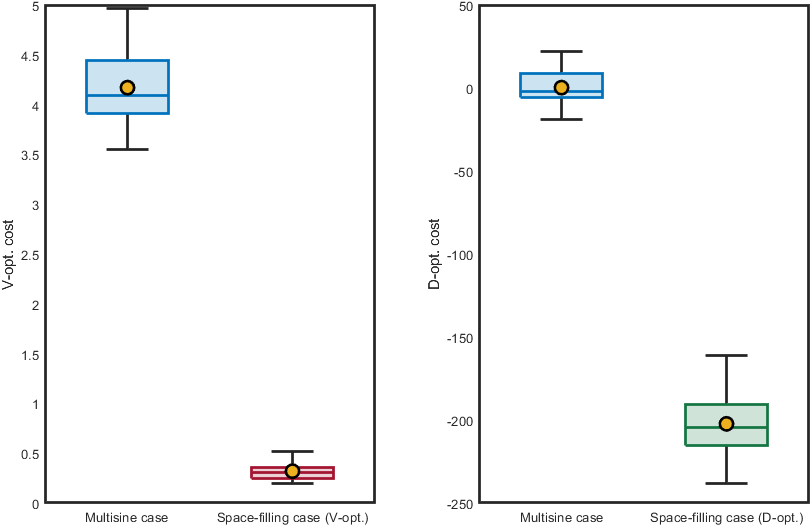}
    \caption{Box plot of the costs over the Monte Carlo simulation. V-optimal case: multisine input (blue) with mean (yellow circle) $4.17$, V-optimal space-filling signal with mean $0.31$ (red). D-optimal case: multisine input (blue) with mean (yellow circle) $0.60$, D-optimal space-filling signal with mean $-201.78$ (green).
    }
    \label{fig:CostboxPlot}
\end{figure}

Figure \ref{fig:signalComparison} shows one example for both V- and D-optimal designs of what the optimized signal \MK{$u_{k,\hat{\theta}}$} looks like compared to the initial multisine \MK{$u_{k,\theta}$}.
Comparing the costs of the different optimality criteria in Figure \ref{fig:CostboxPlot}, the Shannon entropy of the estimator (D-optimal cost) is 332 times lower on average (0.60 vs. -201.78), and the average posterior variance (V-optimal cost) with the space-filling signal is 13 times lower (4.17 vs. 0.31) than the cost obtained with the initial random phase multisine.

\begin{figure}%[htb]
    \centering
    \includegraphics[width=1.0\columnwidth]{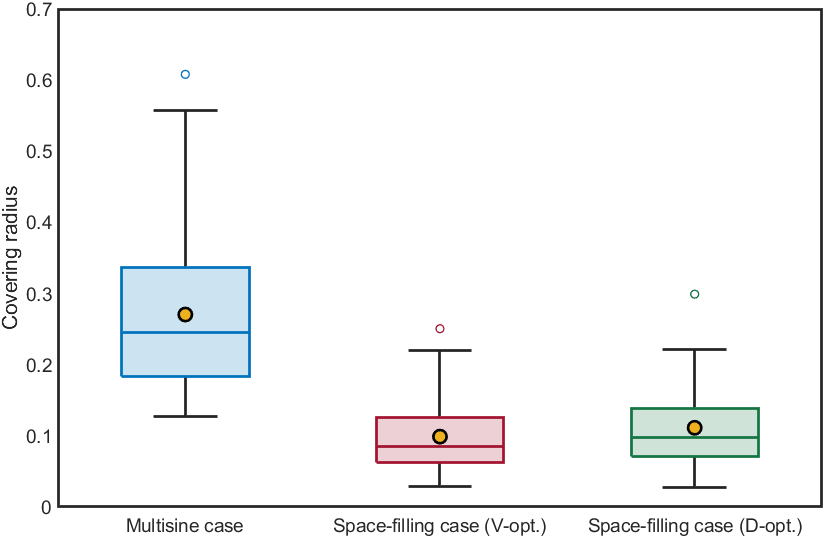}
    \caption{Covering radius box plots of the multisine input (blue) with mean (yellow circle) $0.27$, of the V-optimal space-filling signal with mean $0.09$ (red) and of the D-optimal space-filling signal with mean $0.11$ (green).
    }
    \label{fig:CRboxPlot}
\end{figure}

\begin{figure}%[htb]
    \centering
    \includegraphics[width=0.95\columnwidth]{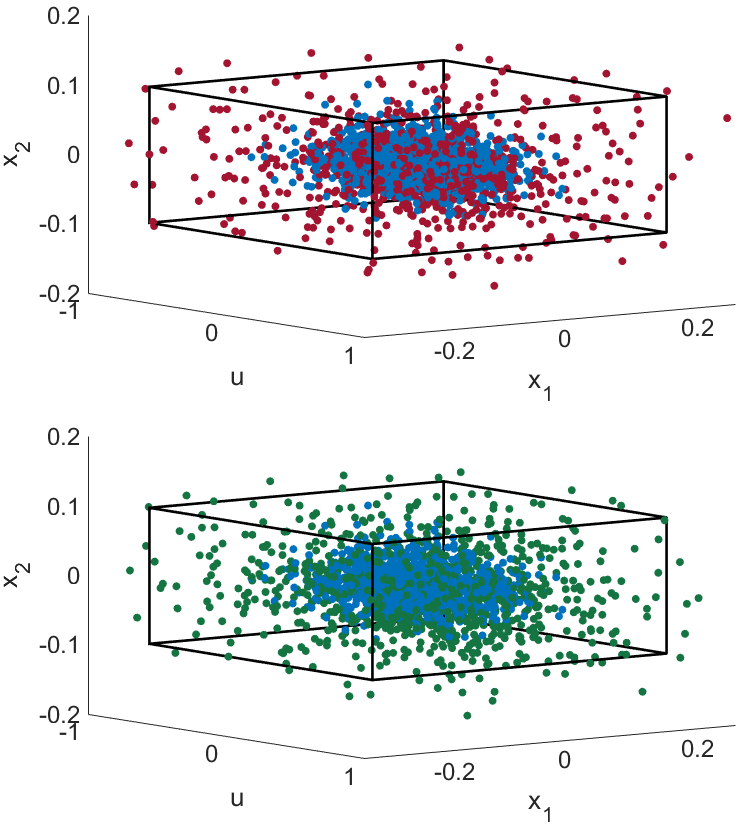}
    \caption{Comparing the initial dataset $\mathcal{D}_N(\theta)$ (blue), V-optimal space-filling data set $\mathcal{D}_{N}^V(\hat{\theta})$ (red) and D-optimal space-filling data set $\mathcal{D}_{N}^D(\hat{\theta})$ (green) in the 3D space for one of the Monte-Carlo runs. The black contour lines outline the borders of the region of interest.
    }
    \label{fig:3D}
\end{figure}

The space-filling performance of the V-optimal and D-optimal data sets is nearly equivalent on average and the coverage increased by a factor of 3.00 in the V-optimal case and by a factor of 2.45 in the D-optimal case (Fig. \ref{fig:CRboxPlot}).
An isometric view of the space-filling effect of the optimized data set is presented in Figure \ref{fig:3D}, showing that the data set coming from the initial input (blue points) is strongly clustered in the middle of the domain, while a clear space-filling effect can be observed after optimization (red and green points).

Note that varying input parametrization can lead to different space-filling behavior. Moreover, using more anchor points further enhances space-filling performance, as demonstrated in \cite{liu2025space}. This section presented the raw observations arising purely from the space-filling behavior of the dataset. Next, we show what benefits this results in for identifying a model. 

\subsection{System Identification}
For each Monte Carlo realization, we identify three models: one with the initial multisine and two other with the V-optimal and D-optimal space-filling signal. Since the space-filling signal captures more of the nonlinear behavior of the model, the identification method must be capable of appropriately handling this complexity. Such a difficulty can be overcome by identifying a black-box nonlinear state space model, using similar model parametrization as \eqref{eq:polyNonlinSS}:
\begin{equation}
    \begin{aligned}
    \xkp &= A\xk + B\uk + \tilde{g}_{\eta}(\xk,\uk), \\
    \yk &= C\xk + D\uk.
    \end{aligned}
\end{equation}
where $\tilde{g}_{\eta}(\xk, \uk)$ is a parametrized as a neural network.
The estimation of the state-space neural network (SS-NN) is initialized by the best linear approximation \cite{schoukens2016linear}, that gives a good linear approximate model of the nonlinear system, and the nonlinearity $\tilde{g}_{\eta}$ is represented by a feedforward neural network:
\begin{equation}
    \tilde{g}_{\eta}(\xk, \uk)\!=\!W_x \sigma \left( \begin{bmatrix} W_{fx} & W_{fu} \end{bmatrix} \begin{bmatrix} \xk \\ \uk \end{bmatrix}\!+\!b_f \right)\!+\!b_x
\end{equation}
with $\sigma$ being a hyperbolic tangent activation function, $W_x,W_{fx},W_{fu}$ representing the weights of the neural network while $b_f$, $b_x$ are the biases. In our system identification setup, we use one hidden layer of 15 neurons and a linear output layer.
The model parameters are estimated through the minimization of the mean squared simulation error:
\begin{equation}\label{eq:SS-NN cost}
V_N(\eta) = \frac{1}{N} \sum_{k=1}^{N} (\yk - \hat{y}(k|\eta))^2,
\end{equation}
\vspace{-5mm}
\begin{equation}
\hat{\eta} = \arg \min_{\eta} V_N(\eta),
\end{equation}
where $\hat{y}(k|\eta)$ is the simulated output of the SS-NN model given the parameter vector $\eta$, the noisy system output $\yk$ and $N$ the total number of samples. The parameter vector $\eta$ contains all the model parameters of the neural network (i.e. weight, biases, state-space matrix entries). The cost function \eqref{eq:SS-NN cost} is minimized with a \emph{Levenberg-Marquardt} algorithm. We refer to \cite{schoukens2021improved} for further reading on the used identification approach.

\begin{figure}%[htb]
    \centering
    \includegraphics[width=0.95\columnwidth]{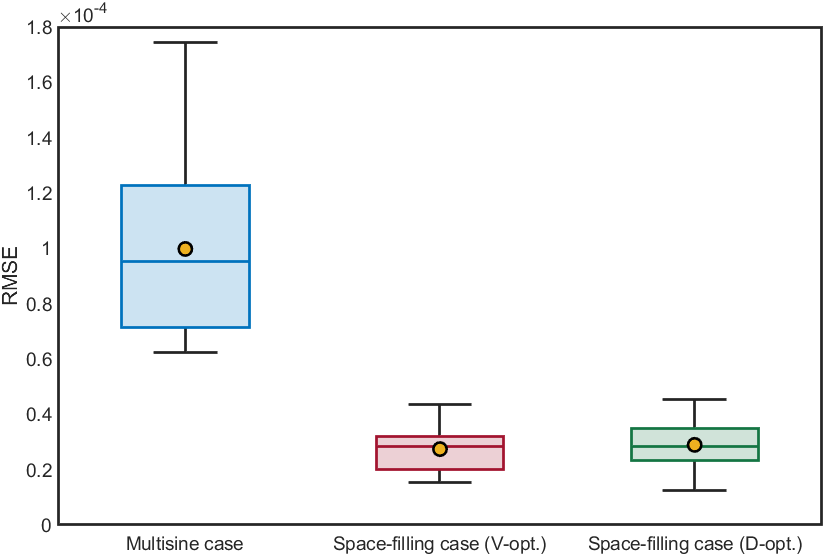}
    \caption{Validation results based on multisine test signal. Initial signal (blue) with mean (yellow circle) $9.97 e-4$, V-optimal space-filling signal (red) with mean $0.28e-4$ and D-optimal space-filling signal (green) with mean $0.28e-4$ of the RMSE errors.
    }
    \label{fig:ValidationOnMultisine}
\end{figure}

As a measure of model quality, the simulation RMSE (root mean square error) is used to compare the results:
\begin{equation}
    e_{\text{RMSE}} = \sqrt{\frac{1}{N} \sum_{k=1}^{N} (\yk - \hat{y}(k|\eta))^2},
    \label{eq:RMSE}
\end{equation}
where $\yk$ is the observed noise free test output and $\hat{y}(k|\eta)$ is the simulated output using the estimated model. To validate our results we use two different test set. The first test data set is generated by a random phase multisine. It excites the same frequency interval using a flat frequency spectrum and the same standard deviation as the initial unoptimized random phase multisine signal except that the number of data samples in one period is $N_{test}=2^{16}$ (i.e. meaning that the test signal is coming from the same input class as the initial multisine). Figure~\ref{fig:ValidationOnMultisine} gives the summarizing overview of the model errors coming from the 50 realizations. The space-filling data sets are superior compared to the initial multisine one even though that the test signal shares the same power spectrum with the initial multisine signal.

\begin{figure}%[htb]
    \centering
    \includegraphics[width=0.95\columnwidth]{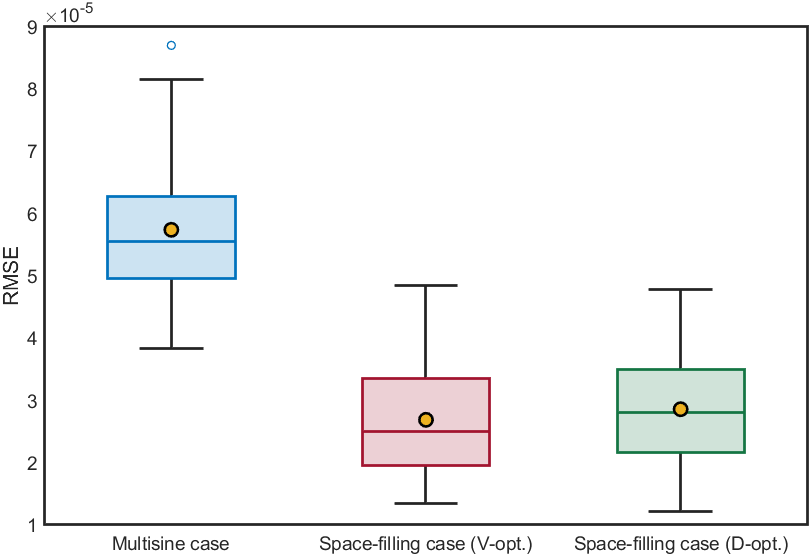}
    \caption{Validation results based on Schroeder test signal. Initial signal (blue) with mean (yellow circle) $5.74 e-5$, V-optimal space-filling signal (red) with mean $2.68e-5$ and D-optimal space-filling signal (green) with mean $2.85e-5$ of the RMSE errors.
    }
    \label{fig:ValidationOnSchroeder}
\end{figure}

\begin{figure}%[htb]
    \centering
    \includegraphics[width=1.0\columnwidth]{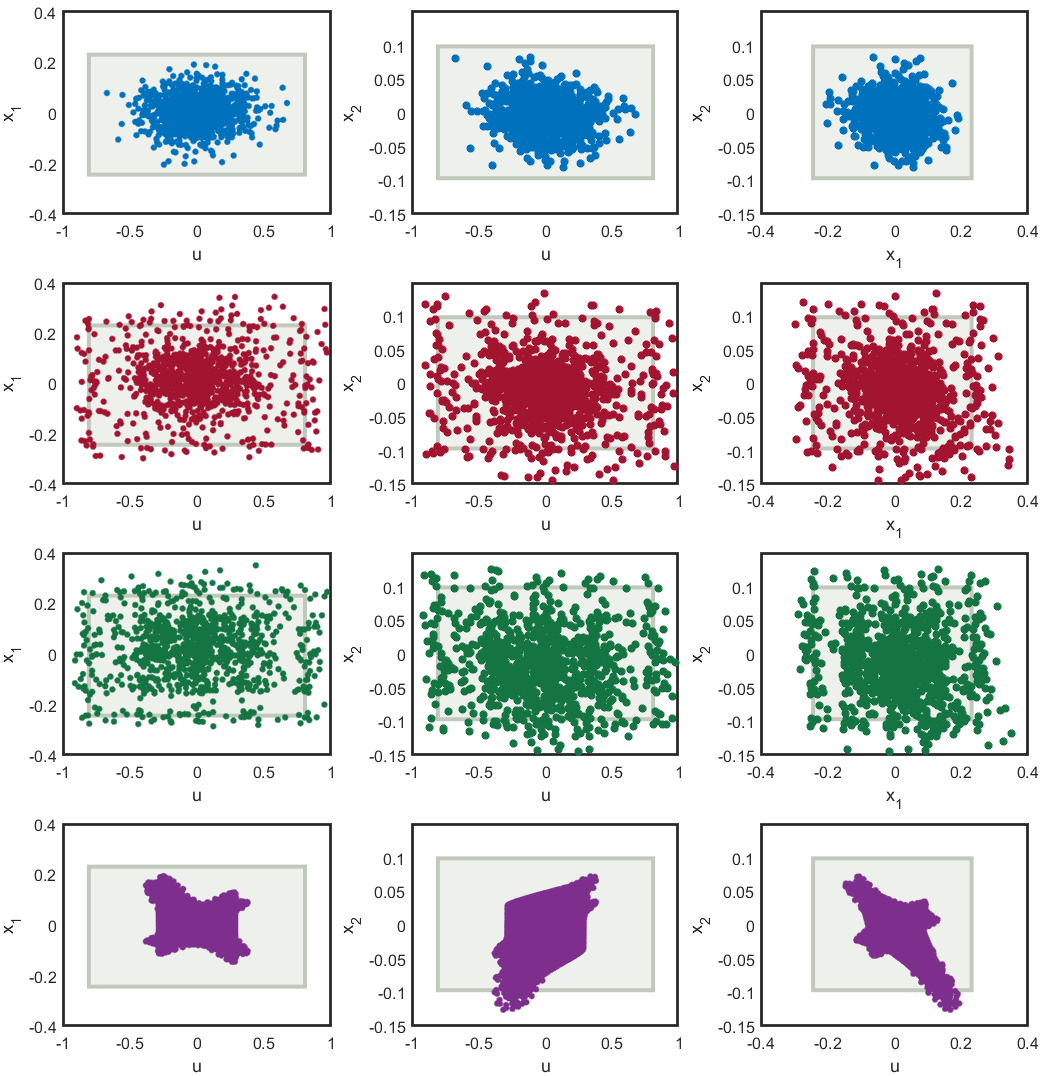}
    \caption{Side views of the space-filling effect. Multisine (blue), V-optimal (red), D-optimal (green) space-filling signal and Schroeder sine (purple). The gray shaded are denotes the region of interest.
    }
    \label{fig:SideViews}
\end{figure}

Continuing the analysis, a second test data set is created with a Schroeder multisine. The signal is constructed by \eqref{eq:multisineInit} using the same attributes introduced for the unoptimized multisine with the exception of the increased data samples $N_{test}=2^{16}$ and phases. Each phase is constructed according to Schroeder`s equation:
\begin{align}
    \varphi_l = \frac{-l(l-1) \pi}{N_l} ,
\end{align}
where $N_l$ is the number of excited frequencies. Figure~\ref{fig:SideViews} shows that the distribution of the Schroeder data set overlaps more with the multisine data set. This overlap, along with the shared amplitude spectrum between the unoptimized multisine and Schroeder signal, suggests that models identified using the multisine signal should perform better at all. Indeed, there are cases in the lower quartile with smaller model errors. However, the space-filling behavior shapes the model errors such that they are relatively small everywhere over the region of interest. Due to this unique property, the models generalize better and the achieved simulation RMSE is lower in average, showed in Figure~\ref{fig:ValidationOnSchroeder}. Namely, the worst performing model identified with a space-filling input signal has lower model error than the average of identified with an initial multisine.

The average model quality of the V-optimal and D-optimal data sets is similar in both cases. 
The difference between them primarily comes from the meaning of the different optimality criteria (see Sec.~\ref{sec:BayesianOptimalSpaceFillingInputDesign}). This shows that the chosen optimality criteria are equally appropriate in this particular use case. However, it is worth noting that this may not hold in every experiment. Since different criteria provide different types of information, it is possible to find such a criteria that significantly outperforms an another one \cite{fedorov1972theory}.

\section{Conclusion}\label{sec:Conclusion}
We showed that the Gaussian Process based space-covering approach connects to the field of optimal experiment design. By formulating the utility function as a functional of the posterior covariance matrix, the Bayesian D- and V-optimality criteria is used to design space-filling experiments and as a result, the optimized signal directly carries the information value of the experiment. Theoretical analysis demonstrates the creation of a certain alphabetic-optimal type of space-filling cost function and the equivalence between the convergence of the optimal cost functions and the expected space covering behavior.\\
In nonlinear system identification, the true system is rarely part of the model set. As a result, the estimated model inherently has a structural error, which highly depends on the input used during identification. In the Monte Carlo study, we demonstrated that the value of the space-covering input design, lies in its ability to shape the model error in a way that ensures an overall good performance over the range of operation. The models identified by the optimal space-covering signals generalize better even for signals which are actually belonging to the same input class as the one that we would have used for the non-optimized signal.
%\vspace{-0.26cm}

\bibliographystyle{plain}        % Include this if you use bibtex 
\bibliography{autosam}           % and a bib file to produce the 
%\cite{schoukens2016linear} % Example citation to fix the issue
                                 % bibliography (preferred). The
                                 % correct style is generated by
                                 % Elsevier at the time of printing.

%\begin{thebibliography}{99}     % Otherwise use the  
                                 % thebibliography environment.
                                 % Insert the full references here.
                                 % See a recent issue of Automatica 
                                 % for the style.
%  \bibitem[Heritage, 1992]{Heritage:92}
%     (1992) {\it The American Heritage. 
%     Dictionary of the American Language.}
%     Houghton Mifflin Company.
%  \bibitem[Able, 1956]{Abl:56}
%     B.~C.~Able (1956). Nucleic acid content of macroscope. 
%     {\it Nature 2}, 7--9. 
%  \bibitem[Able {\em et al.}, 1954]{AbTaRu:54}   
%     B.~C. Able, R.~A. Tagg, and M.~Rush (1954).
%     Enzyme-catalyzed cellular transanimations.
%     In A.~F.~Round, editor, 
%     {\it Advances in Enzymology Vol. 2} (125--247). 
%     New York, Academic Press.
%  \bibitem[R.~Keohane, 1958]{Keo:58}
%     R.~Keohane (1958).
%     {\it Power and Interdependence: 
%     World Politics in Transition.}
%     Boston, Little, Brown \& Co.
%  \bibitem[Powers, 1985]{Pow:85}
%     T.~Powers (1985).
%     Is there a way out?
%     {\it Harpers, June 1985}, 35--47.

%\end{thebibliography}

\appendix
  
                                        % in the appendices.
\end{document}